\documentclass[11pt,journal,twoside,onecolumn]{IEEEtran}
\pdfoutput=1
\usepackage[margin=2.3cm,bottom=2.5cm]{geometry}

\usepackage[ansinew]{inputenc}
\usepackage{color}
\usepackage{colortbl,xcolor}
\usepackage{array}
\usepackage{amsmath}
\usepackage{amssymb}
\usepackage{float}
\usepackage{amscd}
\usepackage{ntheorem}
\usepackage{amsfonts}
\usepackage{amssymb}
\usepackage{cite}
\usepackage{psfrag}
\usepackage[dvips]{graphicx}
\usepackage{booktabs}
\usepackage{pdfcomment}
\definecolor{Gray}{gray}{0.95}

\usepackage[caption=false, font=footnotesize]{subfig}

\newcommand{\w}{{\mathbf w}}
\newcommand{\wo}{{\mathbf w}_\text{o}}

\newcommand{\e}{{\mathbf e}}
\newcommand{\uu}{{\mathbf u}}

\newcommand{\tr}{\mbox{Tr}}

\newcommand{\Ibf}{\mathbf{I}}

\newcommand{\Qbf}{\mathbf{Q}}
\newcommand{\qbf}{\mathbf{q}}
\newcommand{\Rbf}{\mathbf{R}}

\newcommand{\sbf}{\mathbf{s}}

\newcommand{\wbf}{\mathbf{w}}

\definecolor{mygreen}{rgb}{0.7,0.9,0.7}
\definecolor{myblue}{rgb}{0.75,0.94,1}

\title{Combinations of Adaptive Filters}
\author{Jerónimo Arenas-García,~\IEEEmembership{Senior Member,~IEEE}, Luis A.~Azpicueta-Ruiz,~\IEEEmembership{Member,~IEEE},\\  Magno T.~M.~Silva,~\IEEEmembership{Member,~IEEE}, Vítor H.~Nascimento,~\IEEEmembership{Senior Member,~IEEE}, Ali~H.~Sayed,~\IEEEmembership{Fellow, IEEE}
}

\begin{document}

\maketitle
%
%
\IEEEpeerreviewmaketitle

\par Adaptive filters are at the core of many signal processing applications, ranging from acoustic noise supression to echo cancellation \cite{Hansler06}, array beamforming \cite{Widrow01}, channel equalization \cite{Johnson98}, and to  more recent sensor network applications in surveillance, target localization and tracking. A trending approach in this direction is to recur to in-network distributed processing in which individual nodes implement adaptation rules and diffuse their estimation to the network \cite{Sayed14,Sayed14b}.

\par Ranging from the simple Least-Mean-Squares (LMS) to sophisticated state-space algorithms, significant research has been carried out over the last 50 years to develop effective adaptive algorithms, in an attempt to improve their general properties in terms of convergence, tracking ability, steady-state misadjustment, robustness, or computational cost \cite{Sayed08}. Many design procedures and theoretical models have been developed, and many novel adaptive structures are continually proposed with the objective of improving filter behavior with respect to well-known performance tradeoffs (such as convergence rate {\em versus} steady-state performance), or incorporating available {\em a priori} knowledge into the learning mechanisms of the filters (e.g., to enforce sparsity).

\par When the {\em a priori} knowledge about the filtering scenario is limited or imprecise, selecting the most adequate filter structure and adjusting its parameters becomes a challenging task, and erroneous choices can lead to inadequate performance. To address this difficulty, one useful approach is to rely on combinations of adaptive structures.  Combinations of adaptive schemes have been studied in several works \cite{Andersson85,Niedzwiecki90,Singer99,Kozat00,Manel02,Arenas06,Zhang06,Arenas05,Silva08,Bershad08,Bermudez11,Candido10,Kozat10, Kozat11,Trump10} and have been applied to a variety of applications such as the characterization of signal modality \cite{Jelfs10}, acoustic echo cancellation \cite{Arenas09,Ni10,Azpicueta11,Azpicueta13}, adaptive line enhancement \cite{Trump11}, array beamforming \cite{Lu12,Comminiello13}, and active noise control \cite{Ferrer13,George14}. 

\par The combination of adaptive filters exploits to some extent the same {\em divide and conquer} principle that has also been successfully exploited by the machine learning community (e.g., in bagging or boosting \cite{Kuncheva04}). In particular, the problem of combining the outputs of several learning algorithms ({\em mixture-of-experts}) has been studied in the computational learning field under a different perspective: rather than studying the expected performance of the mixture, deterministic bounds are derived that apply to individual sequences and, therefore, reflect worst-case scenarios \cite{CesaBianchi97,Vovk01,Ozkan15}. These bounds require assumptions different from the ones typically used in adaptive filtering, which is the emphasis of this overview article. In this work, we review the key ideas and principles behind these combination schemes, with emphasis on design rules. We also illustrate their performance with a variety of examples.

\section{Problem formulation and notation}
We generally assume the availability of a reference signal and an input regressor vector, $d(n)$ and $\uu(n)$, respectively, which satisfy a linear regression model of the form:
\begin{equation}
\label{eq:linear_reg_model}
d(n) = \uu^\top(n) \wo(n) + v(n),
\end{equation}
where $\wo(n)$ represents the (possibly) time varying optimal solution, and $v(n)$ is a noise sequence, which is considered i.i.d., and independent of $\uu(m)$, for all $n,m$. In this paper, we will mostly restrict ourselves to the case in which all involved variables are real, although the extension to the complex case is straightforward, and has been analyzed in other works \cite{Sayed08}. In order to estimate the optimal solution at time $n$, adaptive filters typically implement a recursion of the form
$$\w(n+1) = \mathbf{f} \left[\w(n), d(n), \uu(n), \sbf(n) \right],$$
where different adaptive schemes are characterized by their update functions $\mathbf{f}[\cdot]$, and $\sbf(n)$ represents any other state information that is needed for the update of the filter.

We define the following error variables that are commonly used to characterize the performance of adaptive filters \cite{Sayed08}:
\begin{itemize}
\item Filter output: $y(n) = \uu^\top(n) \w(n)$
\item Weight error: $\widetilde{\wbf}(n) = \wo(n) - \wbf(n)$
\item {\em A priori} filter error: $e_a(n) = \uu^\top(n) \widetilde{\wbf}(n)$
\item Filter error: $e(n) = d(n) - \uu^\top(n) \w(n) = e_a(n) + v(n)$
\item Mean Square Error: $\text{MSE}(n) = \mathbb{E}\{e^2(n)\}$
\item Excess MSE (EMSE): $\zeta(n) = \mathbb{E}\{e_a^2(n)\} = \text{MSE}(n) - \mathbb{E}\{v^2(n)\}$
\item Mean Square Deviation: $\text{MSD}(n) = \mathbb{E}\{\|\widetilde{\wbf}(n)\|^2\}$
\end{itemize}
During their operation, adaptive filters normally go from a convergence phase, where the expected mean-square error decreases, to a steady-state regime in which the mean-square error tends towards some asymptotic value. Thus, for steady-state performance we also define the steady-state MSE, EMSE, and MSD as their limiting values as $n$ increases. For instance, the steady-state EMSE is defined as $$\zeta(\infty) = \lim\limits_{n\to\infty} \mathbb{E}\{e_a^2(n)\}.$$

\section{A basic combination of two adaptive filters}
\par
\begin{figure}[h]
\centerline{\includegraphics[width=.75\columnwidth,trim= 5cm 6.5cm 4.2cm 5.5cm,clip=true]{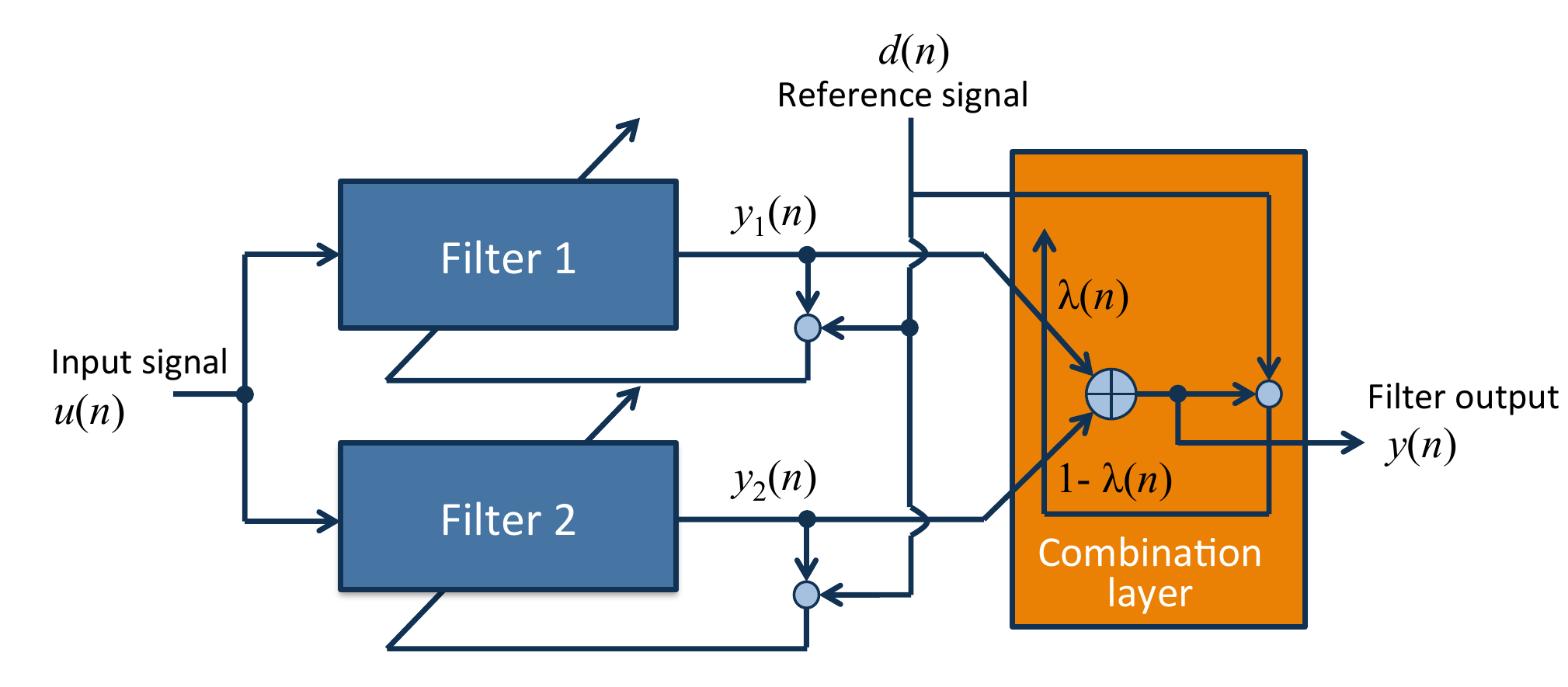}}
\caption{Basic scheme for the combination of two adaptive filters.\label{fig:combination_scheme}}
\end{figure}

 The most simple combination scheme incorporates two adaptive filters. Figure \ref{fig:combination_scheme} illustrates the configuration, which shows that combination schemes have two concurrent adaptive layers: adaptation of individual filters and adaptation of the combination layer. As illustrated in the figure, both adaptive filters have access to the same input and reference signals and produce their individual estimates of the optimum weight vector, $\wo(n)$. The goal of the combination layer is to learn which filter component is performing better dynamically at any particular time, assigning them weights to optimize the overall performance.


In this paper, we focus on affine and convex combinations, where the overall filter output is given by
\begin{equation}
\label{eq:combi_y}
y(n) = \lambda(n) y_1(n) + [1 - \lambda(n)] y_2(n),
\end{equation}
where $y_i(n) = \uu^\top(n) \w_i(n),~i=1,2,$ are the outputs ot the two adaptive filters characterized by weights $\w_i(n)$, and $\lambda(n)$ is a mixing parameter. Similarly, the estimated weight vector, the error, and the {\em a priori} error of the combination scheme are given by
\begin{align}
\w(n) & = \lambda(n) \w_1(n) + [1 - \lambda(n)] \w_2(n), \label{eq:w_comb} \\
e(n) & = \lambda(n) e_1(n) + [1 - \lambda(n)] e_2(n), \label{eq:err_combination}\\
e_a(n) & = \lambda(n) e_{a,1}(n) + [1 - \lambda(n)] e_{a,2}(n),
\end{align}
where $e_i(n) = d(n) - y_i(n),~i=1,2$, are the errors of the two adaptive filter components, and $e_{a,i}(n)$ their {\em a priori} errors. The combination parameter $\lambda(n)$ is constrained to lie in the interval $[0,1]$ for convex combinations, while it can take any real value for affine combinations. There are also combinations where the sum of weights assigned to the component filters is not equal to one \cite{Kozat00}. These unconstrained combinations will not be addressed in this paper. In order to obtain an enhanced behavior from the combination, it is crucial to design mechanisms to learn the mixing parameter $\lambda(n)$ and to adapt it to track drifts in the model. Different algorithms for the adaptation of the combination have been proposed in the literature; some of them will be reviewed in a later section.

In \eqref{eq:w_comb}, we are implicitly assuming component filters with the same length. When this is not the case, we can still use \eqref{eq:w_comb} by extending the shortest filter and filling with zeros the nonexistent coefficients. Using this observation, most of the algorithms discussed in this paper continue to be valid if the component filters have different lengths. Moreover, in practice, construction \eqref{eq:w_comb} does not need to be evaluated since only the output combination \eqref{eq:combi_y} is necessary. Furthermore, although not required by the analysis, in the experiments we will assume that the component filter weight vectors match the length of the optimum solution, $\wo(n)$. Nevertheless, there have been useful studies in the literature where the length of the filters is treated as a design variable and the adaptation rules are used to learn the length of the optimum solution as well to avoid degradation due to under-modeling or over-modeling \cite{Singer99,Zhang06,Kozat10}.

The selection of the individual filters is often influenced by the application itself. For example, combination structures can be used to:
\begin{itemize}
\item Facilitate the selection of filter parameters (such as the step size or forgetting factor, regularization constants, filter length, projection order, etc). Existing schemes fix the values of these parameters or recur to complex schemes that learn them over time. Alternatively, we can consider a combination structure where the two filters belong to the same family of algorithms, but differ in the values of the filter parameters. The combination layer would then select and give more weight to the best filter at every time instant, making parameter selection less critical.
\item Increase robustness against unknown environmental conditions. A common objective in the design of adaptive filters consists in the incorporation of {\em a priori} knowledge about the filtering scenario to improve the performance of the filter. For instance, sparse-aware adaptive filters try to benefit from the kwnowledge that many coefficients of the optimal solution are (close to) zero, and Volterra filters try to model non-linearities with polynomial kernels. However, when the filtering conditions are not known accurately, or change over time, these schemes can show very suboptimal performance. Combination schemes can  be exploited to increase the robustness of these techniques against lack of or imprecise knowledge of the adaptive filtering scenario.
\item Provide diversity that can be exploited to enhance performance beyond the capabilities of an individual adaptive filter. An example will be given later, when we study the tracking abilities of a combination of one LMS and one recursive least-squares (RLS) filters.
\end{itemize}
The main disadvantage of this approach is the increased computational cost required for the adaptation of the two adaptive filters. However, it should be noted that the filters can be adapted in parallel and some strategies allow for the joint adaptation of the two filters with just a slight increment with respect to that of a single adaptive filter. Regarding the adaptation of the mixing parameter, the computational requirement of most available schemes is not significant.

In the next section, we review the theoretical limits of the combination scheme consisting of just two filters. Then, we review different combination rules that have already appeared in the literature, and compare them in a variety of simulated conditions.

\section{Optimum mixing parameter and combination performance}
\label{sec:optimumlambda}
In this section, we derive the expression for the optimal mixing parameter in the affine and convex cases, in the sense of minimizing the MSE of the combination. Since for $\lambda(n) = 0$ and $\lambda(n)=1$ convex and affine combinations are equivalent to each of the component filters, we know that an optimal selection of the mixing parameter would guarantee that these combinations perform at least as the best component filter. In this section, we examine the behavior of the combination scheme for other non-trivial choices of the mixing parameter.

To start with, it can be easily seen that the EMSE of the combination \eqref{eq:combi_y} is given by
\begin{equation}
\label{eq:EMSE_combi}
\zeta(n) = \mathbb{E}\{e_a^2(n)\} = \lambda^2(n) \zeta_1(n) + [1-\lambda(n)]^2 \zeta_2(n) + 2 \lambda(n) [1-\lambda(n)] \zeta_{12}(n),
\end{equation}
where $\zeta_i(n),~i=1,2$, are the EMSEs of the two individual filters, and $\zeta_{12}(n)$ denotes the cross-EMSE defined as \cite{Arenas06}:
$$\zeta_{12}(n) = \mathbb{E}\{e_{a,1}(n) e_{a,2}(n)\}.$$
This cross-EMSE variable is an important measure and is closely related to the ability of the combination to improve the performance of both filter components, as we will see shortly.
An important property of the cross-EMSE that will be useful when discussing the properties of the optimal combination is derived from the Cauchy-Schwartz inequality, namely:
\begin{equation}
|\zeta_{12}(n)|^2 \leq \zeta_1(n) \zeta_2(n),
\end{equation}
which implies that the cross-EMSE can never be simultaneously larger (in absolute terms) than both individual EMSEs.

Expression \eqref{eq:EMSE_combi} reveals a quadratic dependence on the mixing factor, $\lambda(n)$. It follows that, in the affine case, the optimal choice for $\lambda(n)$ is given by:
\begin{equation}
\label{eq:lambda_opt}
\lambda^\text{o}_{\text{aff}}(n) = \frac{\zeta_2(n) - \zeta_{12}(n)}{\zeta_1(n) + \zeta_2(n) - 2 \zeta_{12}(n)} = \frac{\Delta \zeta_2(n)}{\Delta \zeta_1(n) + \Delta \zeta_2(n)},
\end{equation}
where we have defined $\Delta\zeta_i(n) = \zeta_i(n) - \zeta_{12}(n),~i=1, 2$. In the case of a convex combination, the minimization of \eqref{eq:EMSE_combi} needs to be carried out by constraining $\lambda(n)$ to lie in the interval $[0,1]$. Given that \eqref{eq:EMSE_combi} is non-negative and quadratic in $\lambda(n)$, the optimum mixing parameter is either given by \eqref{eq:lambda_opt} or lies in the limits of the considered interval, i.e., 
\begin{equation}
\label{eq:lambda_cvx}
\lambda^\text{o}_{\text{cvx}}(n) = \frac{\Delta \zeta_2(n)}{\Delta \zeta_1(n) + \Delta \zeta_2(n)}\Bigg\vert_{0}^1,
\end{equation}
where the vertical line denotes truncation to the indicated values.

Substituting either \eqref{eq:lambda_opt} or  \eqref{eq:lambda_cvx} into \eqref{eq:EMSE_combi}, we find that the optimum EMSE of the filter is given by
\begin{subequations}
\begin{align}
\label{eq:EMSE_opti}
\zeta^\text{o}(n) & = \zeta_1(n) - \left[ 1 - \lambda^\text{o}(n)\right]\Delta \zeta_1(n) \\ & = 
 \zeta_2(n) - \lambda^\text{o}(n) \Delta \zeta_2(n),\label{eq:EMSE_opti2}
\end{align}
\end{subequations}
where $\lambda^\text{o}$ stands for either $\lambda^\text{o}_{\text{aff}}(n)$ or $\lambda^\text{o}_{\text{cvx}}(n)$. Note that these expressions are valid for any time instant $n$, and also for the steady-state performance of the filter (i.e., for $n\to\infty$). 

The analysis of the expressions for the optimum mixing parameter and for the combination EMSE allows us to conclude that, depending on the cross-EMSE value, the combination can be performing in one of four possible conditions at every time instant. Taking into account that the denominator of \eqref{eq:lambda_opt} is non-negative, $\Delta \zeta_1(n) + \Delta \zeta_2(n) = \mathbb{E}\{[e_{a,1}(n)-e_{a,2}(n)]^2\}$, Table \ref{tab:three_cases} summarizes the main properties of these four possible cases (the time index is omitted in the first column for the sake of compactness).

\begin{table}[h!]
\caption{Possible operation regimes for the affine and convex combination of two adaptive filters.\label{tab:three_cases}}
\begin{center}
\begin{tabular}{llllll}
\toprule
&$\zeta_{12}(n)$ & $\Delta\zeta_1(n)$ & $\Delta\zeta_2(n)$ & $\lambda^\text{o}_\text{aff}(n)$ & $\lambda^\text{o}_\text{cvx}(n)$\\
\midrule
Case 1: & $\zeta_1 \leq \zeta_{12} ~ (<\sqrt{\zeta_1\zeta_2} <\zeta_2) $ & $\leq0$ & $>0$ & $\geq1$ & $=1$\\
Case 2: & $\zeta_2 \leq \zeta_{12} ~ (<\sqrt{\zeta_1\zeta_2} <\zeta_1) $ & $>0$ & $\leq0$ & $\leq0$ & $=0$\\
Case 3: & $(-\sqrt{\zeta_1\zeta_2} < ) ~ \zeta_{12} < \min\{\zeta_1,\zeta_2\} $ & $>0$ & $>0$ & $\in (0,1)$ & $\in (0,1)$\\
Case 4: & $\zeta_1 = \zeta_2 = \zeta_{12}$ & $=0$ & $=0$ & --- & ---\\
\bottomrule
\end{tabular}
\end{center}
\end{table}

From the information included in the table we can deduce the following properties:
\begin{itemize}
\item In cases 1 and 2, the convexity constraints are active. As a consequence, the optimum mixing parameter of the convex combination is either 1 or 0, respectively, making this scheme behave just like the best filter component. However, an affine combination can outperform both components in this case. To see this, consider \eqref{eq:EMSE_opti} and \eqref{eq:EMSE_opti2} for cases 1 and 2, respectively, and note that a positive value is subtracted from the smallest filter EMSE ($\zeta_1$ and $\zeta_2$) to obtain the EMSE of the combination.
\item In case 3, affine and convex combinations share the same optimum mixing parameter. As in the previous case, using \eqref{eq:EMSE_opti} and \eqref{eq:EMSE_opti2} we can easily conclude that, in this case, combinations outperform both component filters simultaneously. This performance can be explained from the small cross-EMSE in this case: since the correlation between the {\em a priori} errors of both component filters is small, their weighted combination provides an estimation error of reduced variance.
\item For completeness, we have included a fourth case where $\zeta_1 = \zeta_2 = \zeta_{12}$. Simplifying all terms that involve $\lambda(n)$ in \eqref{eq:EMSE_combi}, we can conclude that in this case we have $\zeta^\text{o}(n) = \zeta_1(n) = \zeta_2(n)$, irrespectively of the value of the mixing parameter. It is important to remark that for case 4 to hold, condition $\zeta_1 = \zeta_2 = \zeta_{12}$ should be satisfied exactly, so that this case will be rarely encountered in practice.
\end{itemize}
We should emphasize that the results in this section refer to the theoretical performance of optimally adjusted affine and convex combinations. Practical algorithms to adjust the mixing parameter will be presented in a later section, and will imply the addition of a certain amount of noise that, in many cases, justify the adoption of convexity constraints.



\subsection{Cross-EMSE of LMS and RLS filters}
The cross-EMSE plays an important role in clarifying the performance regime of a combination of two adaptive filters. Several works have extended the available theoretical models for the EMSE of adaptive filters to the derivation of expressions for the cross-EMSE 
\cite{Arenas06,Silva08,Bershad08,Sayed08}. In this section, we review some of the main results for LMS and RLS filters. 

Assume that the optimal solution satisfies the following random-walk model\footnote{An inconvenience of this model is that it implies divergence of the optimum solution, i.e., $\lim_{n\to\infty} \mathbb{E}\{\|\wo(n)\|^2\} = \infty$. Nevertheless, this model has been extensively used in the adaptive filtering literature, as it is simpler and leads to similar results than other more realistic tracking models ---see \cite{Sayed08} for further explanations on this issue.}
\begin{equation}
\label{eq:randomwalk}
\wo(n) = \wo(n-1) + \qbf(n),
\end{equation}
where $\qbf(n)$ is the change of the solution at every step that is assumed to have zero-mean and covariance $\Qbf = \mathbb{E}\{\qbf(n)\qbf^\top(n)\}$. It is known \cite{Sayed08} that in this case there is a tradeoff between filter performance and tracking ability.
The following table shows an approximation of the steady-state EMSE of LMS and RLS adaptive filters derived using the energy conservation approach described in \cite{Sayed08}. From these expressions, it is straightforward to obtain the optimum step size for the LMS adaptation rule ($\mu_\text{o}$) and the optimal forgetting factor for the RLS adaptive filter with exponentially decaying memory ($1-\beta_\text{o}$). These expressions are also indicated in the table, together with the associated optimum EMSE for each filter family, $\zeta_\text{o}(\infty)$.

\begin{table}[h!]
\caption{\label{tab:EMSE}Tracking steady-state performance of LMS and RLS adaptive filters. The table shows steady-state EMSE as a function of the LMS step size ($\mu$) and RLS forgetting factor ($1-\beta$), the optimal values of these parameters, and the minimum EMSE that can be achieved by both adaptive filters [$\zeta_\text{o}(\infty)$].}
\begin{center}
\begin{tabular}{llll}
\toprule
Algorithm & $\zeta(\infty)$ & $\mu_\text{o}$ or $\beta_\text{o}$ & $\zeta_\text{o}(\infty) $\\
\midrule
LMS & $\frac{1}{2} \left[\mu \sigma_v^2 \tr\{\Rbf\} + \mu^{-1} \tr\{\Qbf\}\right] $ & $\sqrt{\frac{\tr\{\Qbf\}}{\sigma_v^2 \tr\{\Rbf\}}}$ & $\sqrt{\sigma_v^2 \tr\{\Rbf\} \tr\{\Qbf\}}$ \\
RLS & $\frac{1}{2} \left[\beta \sigma_v^2 M + \beta^{-1} \tr\{\Qbf\Rbf\}\right] $  & $\sqrt{\frac{\tr\{\Qbf \Rbf\}}{\sigma_v^2 M}}$ & $\sqrt{\sigma_v^2 M \tr\{\Qbf\Rbf\}}$\\
\bottomrule
\multicolumn{4}{l}{\scriptsize$\sigma_v^2 = \mathbb{E}\{v^2(n)\}$; $\Qbf = \mathbb{E}\{\qbf(n)\qbf^\top(n)\}$; $\Rbf = \mathbb{E}\{\uu(n)\uu^\top(n)\}$; $M$ is the filter length}
\end{tabular}
\end{center}
\end{table}

In order to study the behavior of combinations of this kind of filters, references \cite{Arenas05,Silva08} derived expressions for the steady-state cross-EMSE, for cases in which the two component filters belong to the same family (LMS or RLS) and also for the combination of one LMS and one RLS filter. These results are summarized in Table \ref{tab:cross_EMSE}, and constitute the basis for the study that we carry out in the rest of this section.
\begin{table}[h!]
	\caption{\label{tab:cross_EMSE}Steady-state cross-EMSE for the LMS and RLS adaptive filter families.}
	\begin{center}
	\begin{tabular}{ll}
		\toprule
		Combination type & steady-state cross-EMSE, $\zeta_{12}(\infty)$\\
		\midrule
		$\mu_1$-LMS and $\mu_2$-LMS & $\displaystyle\frac{\mu_1 \mu_2 \sigma_v^2 \tr\{\Rbf\} + \tr\{ \Qbf\}}{\mu_1 + \mu_2}$\\
		$\beta_1$-RLS and $\beta_2$-RLS & $\displaystyle\frac{\beta_1 \beta_2 \sigma_v^2 M + \tr\{ \Qbf\Rbf\}}{\beta_1 + \beta_2}$\\
		$\mu$-LMS and $\beta$-RLS & $\mu \beta \sigma_v^2 \tr\{ (\mu \Rbf + \beta \Ibf)^{-1} \Rbf\} + \tr\{\Qbf (\mu \Rbf + \beta \Ibf)^{-1} \Rbf \}$ \\
		\bottomrule
				
	\end{tabular}
\end{center}
\end{table}

\subsection{Combination of two LMS filters with different step sizes}
A first example to illustrate the possibilities of combination schemes consists in putting together two filters with different adaptation speeds to obtain a new filter with improved tracking capabilities. Here, we illustrate the idea with a combination of two LMS filters with different step sizes, as was done in \cite{Arenas06}, but the idea can be straightforwardly generalized to other adaptive filter families (see, among others, \cite{Silva08,Kozat10,Trump10,Arenas09,Azpicueta11,Ferrer13}).

Using the tracking model given by \eqref{eq:randomwalk}, in this section we analyze the theoretical steady-state EMSE for varying $\tr\{\Qbf\}$. In the studied scenario, the product $\sigma_v^2 \tr\{\Rbf\}$ remains fixed and is equal to $10^{-2}$. To facilitate the comparison among filters, we will recur to the Normalized Squared Deviation (NSD), which we define as the EMSE excess with respect to the optimum LMS filter for a particular $\tr\{\Qbf\}$, i.e.,
$$\text{NSD}_i(\infty) = \frac{\zeta_i(\infty)}{\zeta_{\text o}(\infty)},~i=1,2, ~~~~\text{and} ~~~~ \text{NSD}(\infty) = \frac{\zeta(\infty)}{\zeta_{\text o}(\infty)},$$
for the component LMS filters and their combination, respectively.

\begin{figure}
	\centerline{\includegraphics[width=.5\columnwidth,trim= 0 0 0 0,clip=true,angle=0]{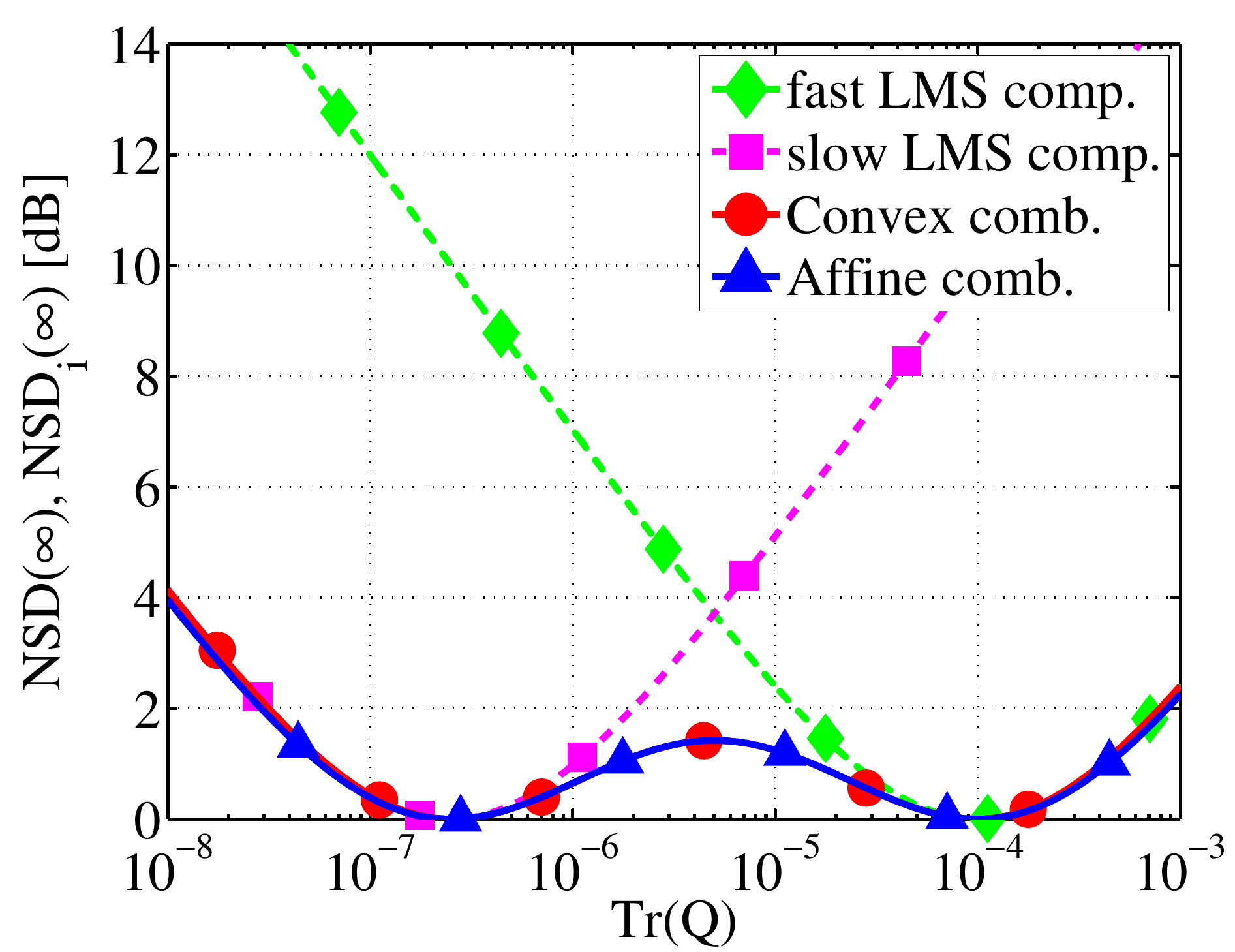}\includegraphics[width=.5\columnwidth,trim= 0 0 0 0,clip=true,angle=0]{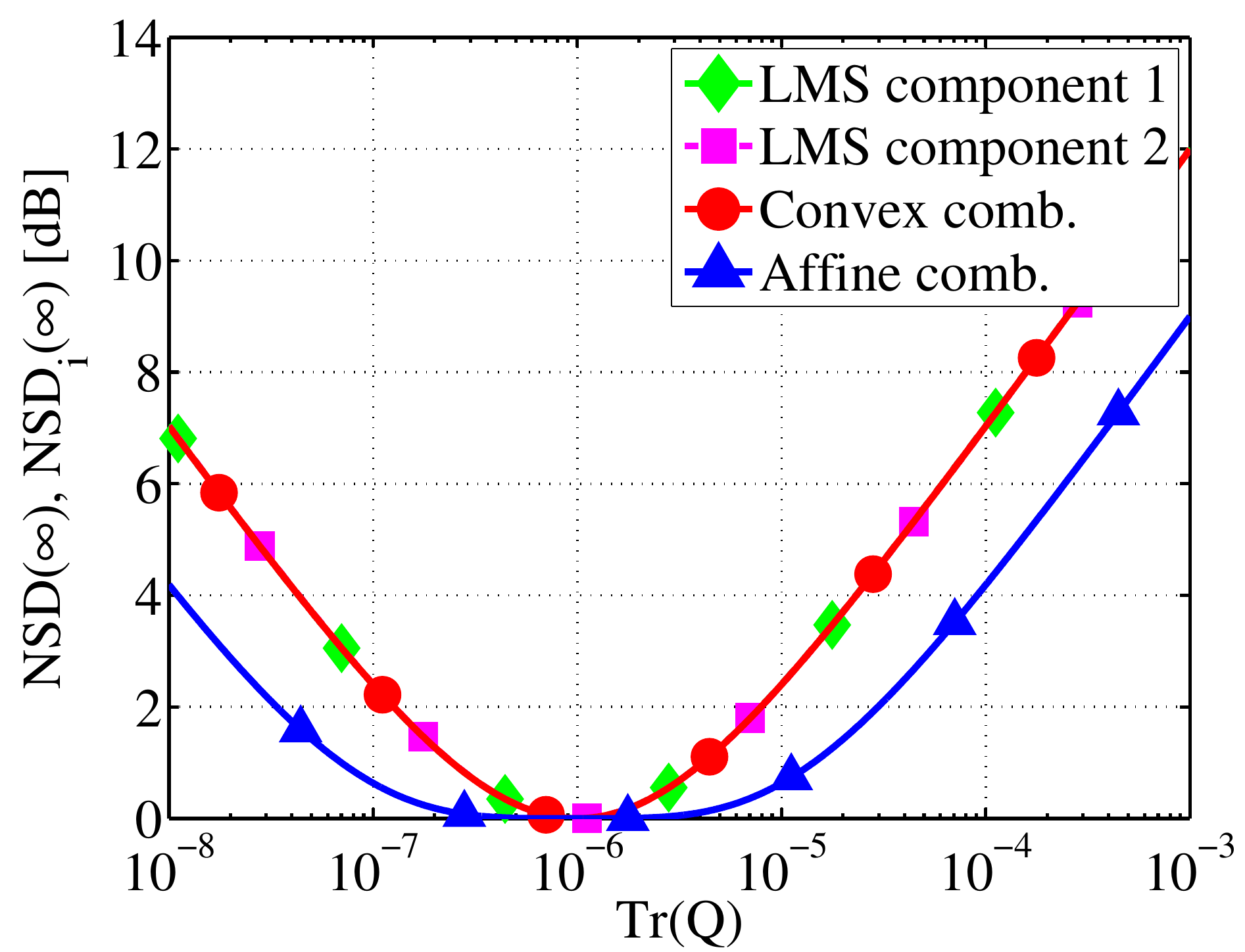}}
	\caption{Steady-state NSD of a combination of two adaptive filters with different step sizes as a function of the speed of changes of the optimum solution. The figures illustrate the NSD incurred by the individual LMS filters, and by their optimum combination. LMS step sizes are: (left) $\mu_1 = 0.1$ and $\mu_2 = 0.005$; (right) $\mu_1 = 0.01$ and $\mu_2 = 0.010001$.\label{LMS_LMS_theo}}
\end{figure}

Figure \ref{LMS_LMS_theo} illustrates the theoretical NSD of  individual LMS filters and combined schemes in two cases. The left panel considers LMS components with step sizes $\mu_1 = 0.1$ and $\mu_2 = \mu_1 / 20$. We can see that both individual filters become optimal (NSD = 0 dB) for a certain value of $\tr\{\Qbf\}$, and very rapidly increase their NSD level both for larger and smaller $\tr\{\Qbf\}$, i.e., for solutions that change relatively faster or slower than the speed of changes for which the filter is optimum. We can see that the NSD of each individual filter remains smaller than 2 dB for a range of about two orders of magnitude of $\tr\{\Qbf\}$. In contrast to this, we can check that the NSD of the combination of the two filters remains smaller than 2 dB when $\tr\{\Qbf\}$ changes by more than four orders of magnitude. It should be mentioned that, in this case, both the affine and convex combinations provide almost identical behavior.

A second possibility is illustrated in the right panel of Figure \ref{LMS_LMS_theo}, for which the step sizes of the two LMS filters are set at almost identical values ($\mu_1 = 0.01$ and $\mu_2 = 0.010001$). In this case, both LMS components obtain approximately the same NSD for all values of $\tr\{\Qbf\}$. According to the theoretical results, it can be shown that in this example cases 1 and 2 enumerated in Table \ref{tab:three_cases} are observed when $\tr({\bf Q})$ is, respectively, smaller or larger than approximately $10^{-6}$. The limitation in the range of $\lambda(n)$ makes the convex combination perform exactly as any of the components, whereas the affine combination achieves an almost systematic gain of 3 dB with respect to the single filters \cite{Candido10}, with the NSD remaining at values of less than 2 dB when $\tr\{\Qbf\}$ is changed by three orders of magnitude. As a result, in this case, the combination approach offers a possibility to improve the tracking capability of LMS filters. The examples in this section illustrate the following points:
\begin{itemize}
\item By employing a combination structure, we can improve the tracking ability of the adaptive implementation. This is important because the actual speed of change of the solution is rarely known {\em a priori}, and can even change over time. Later in Subsection \ref{subsec:vss} we compare the combination framework with variable step-size (VSS) schemes, and show that combinations are more robust than VSS. 
\item When combining two LMS filters with different step sizes, the resulting combination cannot outperform the optimum filter for a given $\tr\{\Qbf\}$ \cite{Silva08}. Therefore, in this situation, if enough statistical information is {\em a priori} available, it would be preferable to use a single LMS filter with optimum step size. A different situation will be described in the next subsection.
\end{itemize}

\subsection{Improving the tracking performance of LMS and RLS filters}
When studying  a combination of LMS and RLS filters following the same approach, an interesting result is obtained. Assume the tracking model \eqref{eq:randomwalk} and consider the optimum LMS and RLS filters with adaptation parameters given by the expressions in Table \ref{tab:EMSE}. As explained in \cite{Sayed08}, LMS will outperform RLS if $\Qbf$ is proportional to the autocorrelation matrix of the input signal, $\Rbf$, and the opposite will occur when $\Qbf \propto \Rbf^{-1}$. To illustrate this behavior, let us consider a synthetic example where $\Qbf$ is a mixture of $\Rbf$ and $\Rbf^{-1}$:
\begin{equation}
\label{eq:Q_model}
\Qbf = 10^{-5} \left[\alpha \frac{\Rbf}{\tr\left\{\Rbf\right\}} + (1-\alpha)  \frac{\Rbf^{-1}}{\tr\left\{\Rbf^{-1}\right\}}\right],
\end{equation}
with $\alpha \in (0,1)$.

\begin{figure}
	\centerline{\includegraphics[width=.8\columnwidth,trim= 2.8cm 4.5cm 4.5cm 5cm,clip=true,angle=0]{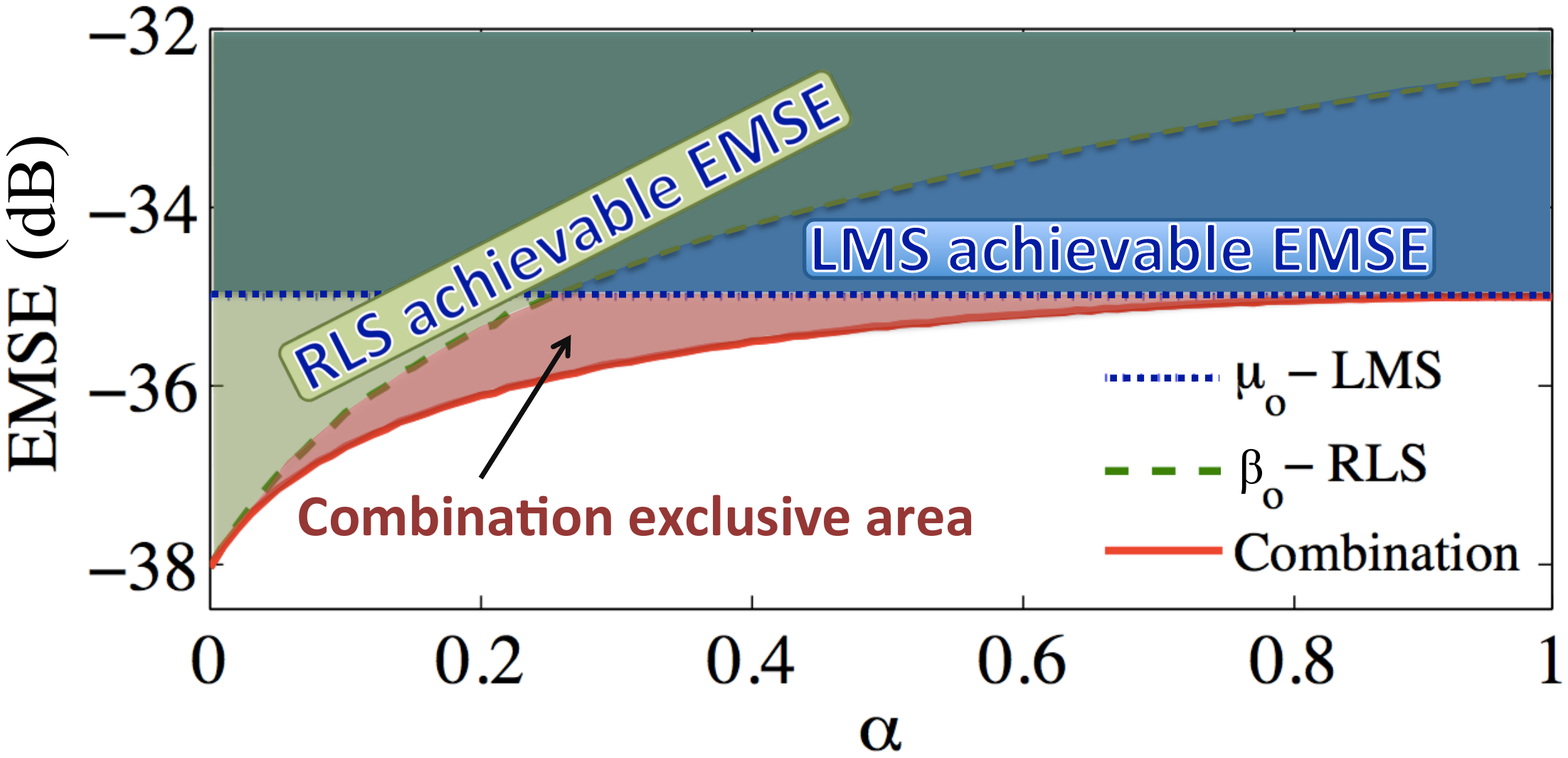}}
	\caption{Tracking performance of a combination of LMS and RLS filters when $\bf Q$ smoothly changes between ${\bf R}^{-1}$ (for $\alpha=0$) and $\bf R$ (for $\alpha = 1$). The performance of optimally adjusted filters is depicted with curves. The blue and light green regions represent EMSEs that can be obtained with LMS and RLS filters, respectively, whereas the dark green area comprises feasible EMSE values for both LMS and RLS filters. Finally, the red region contains EMSE values that can be obtained with combinations of LMS and RLS (but not with these filters individually).\label{RLS_LMS_fig}}
\end{figure}

Figure \ref{RLS_LMS_fig} plots the steady-state optimum EMSE that can be achieved by both types of filters with optimum settings ($\mu_\text{o}$-LMS and $\beta_\text{o}$-RLS), when $\bf Q$ smoothly changes between ${\bf R}^{-1}$ (for $\alpha=0$) and $\bf R$ (for $\alpha = 1$). Other settings for this scenario are: optimal solution and filters length $M=7$, noise variance $\sigma_v^2 = 10^{-2}$, and $\Rbf$ is a Toeplitz matrix whose first row is given by $\frac{1}{7}[1, 0.8, 0.8^2, \cdots, 0.8^{6}]$. According to \eqref{eq:Q_model}, the value of the trace of $\Qbf$ remains constant when changing the value of $\alpha$, meaning that the EMSE of the optimum LMS is approximately equal to $-35$ dB for any value of $\alpha$. If a non-optimum step size is selected, the LMS filter will incur a larger EMSE. Thus, the region of feasible EMSE values for this family of filters is given by the area above the $-35$ dB level in Fig. \ref{RLS_LMS_fig}. Similarly, the region above the curve for the optimum RLS performance represents all feasible EMSE values for the family of RLS filters. We can see that, depending on the value of $\alpha$, the optimal EMSE for RLS filters can be larger or smaller than the optimal EMSE for LMS filters. Finally, the dark green area in Fig. \ref{RLS_LMS_fig} is a feasible region for both LMS and RLS filters.

The performance that can be achieved using a combination of one LMS and one RLS filters is also illustrated. The red curve represents the steady-state tracking EMSE of the combination when using $\mu_\text{o}$ and $\beta_\text{o}$ for the LMS and RLS filters in the combination. As expected, the optimal EMSE for the combination is never larger than the smallest of the filter component EMSEs. More importantly, in this case we can check the existence of an area of EMSE values which is exclusive to the combination scheme. In other words, the red area in Figure \ref{RLS_LMS_fig} is a feasible region for the operation of the combined scheme, but not for single LMS or RLS filters. It can be shown that, in this case, the optimum mixing parameter of the combination lies in interval $(0,1)$, so this conclusion is valid for convex combination schemes \cite{Nascimento10}. This is a useful conclusion that can be used to improve the tracking performance beyond the limits of individual filters.

\section{Estimating the combination parameter}
\label{sec:lambda_learning}
Since the optimum linear combiner \eqref{eq:lambda_opt} is unrealizable,
many practical algorithms have been proposed in the literature to adjust the mixing parameter in convex
\cite{Arenas06,AzpicuetaICASSP08,LazaroTSP10} and affine combinations
\cite{Bershad08,Candido10,Bermudez11,AzpicuetaMLSP08,Trump11journal}.
Rewriting \eqref{eq:combi_y} as
\begin{equation}
y(n) = y_2(n) + \lambda(n) [y_1(n)-y_2(n)],
\end{equation}
we can reinterpret the adaptation of $\lambda(n)$ as a ``second layer'' adaptive filter of length one, so that in principle any adaptive rule can be used for adjusting the mixing parameter. However, this filtering problem has some particularities, namely, the strong and time-varying correlation between $y_1(n)$ and $y_2(n)$. This implies that the power of the difference signal, $[y_1(n)-y_2(n)]$, is also time-varying depending, e.g., on the signal-to-noise conditions and on whether the individual filters are operating in convergence phase or steady state, etc.

Using a stochastic gradient search to minimize the quadratic error $e^2(n)$ of the overall filter, defined in \eqref{eq:err_combination}, reference \cite{Bershad08} proposed the following adaptation for the mixing parameter in an affine combination (i.e., without imposing restrictions on $\lambda(n)$),
\begin{equation}
\lambda(n+1)=\lambda(n)-\frac{\mu_\lambda}{2}\frac{\partial e^2(n)}{\partial \lambda(n)}=\lambda(n)+{\mu}_\lambda e(n)[y_1(n)-y_2(n)].
\end{equation}
We will refer to this case as {\em aff-LMS} adaptation, with ${\mu_\lambda}$ being a step-size parameter. As discussed in \cite{Bershad08}, a large step size should be used in this rule to ensure an adequate behavior of the affine combination. However, this can cause instability during the initial convergence of the algorithm, which was circumvented in \cite{Bershad08} by
constraining $\lambda(n)$ to be less than or equal to one. Even with this constraint, the combination scheme may stay away from the optimum EMSE (see, e.g., \cite{Candido10}). This is a direct consequence of the time-varying power of $[y_1(n)-y_2(n)]$, which makes selection of $\mu_\lambda$ a difficult task.

In order to obtain a more robust scheme, it is possible to recur to normalized adaptation schemes. Using a rough (low-pass filtered) estimation of the power of $[y_1(n)-y_2(n)]$, a power normalized version of {\em aff-LMS} was proposed in \cite{Candido10}, and is summarized in Table \ref{tab:lblearning}. This {\em aff-PN-LMS} algorithm does not impose any constraints on $\lambda(n)$, is less sensitive to the filtering scenario, and converges in the mean-square sense if the step size is selected in the interval $0<{\mu}_{\lambda}<2$.

Rather than directly adjust $\lambda(n)$ as in the affine case, convex combination schemes recur to activation functions to keep the mixing parameter in the range of interest. For example, reference \cite{Arenas06} proposed an adaptation scheme for an auxiliary parameter $a(n)$ that is related to $\lambda(n)$ via the {\em sigmoid} function
\begin{equation}
\lambda(n)={\rm sgm}[a(n)]=\frac{1}{1+{\rm e}^{-a(n)}}.\label{eq:activation}
\end{equation}
Recurring to this activation function (or similar ones), $a(n)$ can be adapted without constraints, and $\lambda(n)$ will be automatically kept inside the interval $(0, 1)$ at all times. Using a gradient descent method to minimize the quadratic error  of the
overall filter, $e^2(n)$, two algorithms were proposed to update $a(n)$: the {\em cvx-LMS} algorithm \cite{Arenas06} and its power normalized version, {\em cvx-PN-LMS} \cite{AzpicuetaICASSP08}, whose update equations are given by
\begin{align}
a(n+1) & = a(n)+\mu_a \lambda(n)[1-\lambda(n)] e(n)[y_1(n)-y_2(n)], ~~~~~~~\text{\em (cvx-LMS)}\label{eq:cvxLMS_nn} \\
a(n+1) & = a(n)+\frac{\mu_a}{p(n)} \lambda(n)[1-\lambda(n)] e(n)[y_1(n)-y_2(n)]. ~~~~\text{\em (cvx-PN-LMS)}\label{eq:cvxLMS}
\end{align}
Here, $p(n)$ is a low-pass filtered estimation of the power of $[y_1(n)-y_2(n)]$.
These algorithms are also shown in Table \ref{tab:lblearning} for further reference. Compared to the {\em cvx-LMS} scheme, {\em cvx-PN-LMS} is more robust to signal-to-noise ratio (SNR)\footnote{According to the linear regression model \eqref{eq:linear_reg_model}, the SNR is defined as $\text{SNR} = [\wo^\top(n) {\bf R} \wo(n)]/{\sigma_v^2}$.} changes, and simplifies the selection of step size $\mu_a$ \cite{AzpicuetaICASSP08}.


\begin{table}[t]
	\centering \caption{\label{tab:lblearning} Power normalized adaptation of $\lambda(n)$
		in affine combinations and of auxiliary parameter $a(n)$ in convex combinations. Unnormalized rules {\em aff-LMS} and {\em cvx-LMS} are obtained setting $p(n)=1$.}
	\begin{center}
		\begin{tabular}{ll}
			\toprule
			Algorithm & Update Equations \\
			\midrule
			{\em aff-PN-LMS}\cite{Candido10}& $\displaystyle \lambda(n+1)=\lambda(n)+\frac{{\mu}_\lambda}{\epsilon+p(n)} e(n)[y_1(n)-y_2(n)]$ \\
			& {\scriptsize $p(n) = \eta~p(n - 1) + (1 - \eta)[y_1(n) - y_2(n)]^2$}\\
			& $0\ll\eta<1$; \;\;\;\; $\epsilon>0$ is a small constant\\
			{\em cvx-PN-LMS} \cite{AzpicuetaICASSP08} 
			& $\displaystyle a(n+1)=a(n)+\frac{\mu_a}{p(n)}\lambda(n)[1-\lambda(n)] e(n)[y_1(n)-y_2(n)]$    \\
			& {\scriptsize $p(n) = \eta~p(n - 1) + (1 - \eta)[y_1(n) - y_2(n)]^2$}\\
			& $0\ll\eta<1$; \;\;\;\; Constraint: $-a^{+}\leq a(n+1)\leq a^{+}$    \\
			\bottomrule
		\end{tabular}
	\end{center}
	\end{table}

The factor $\lambda(n)[1-\lambda(n)]$, that appears in \eqref{eq:cvxLMS}, reduces the gradient noise when $\lambda(n)$ gets too close to the values of zero or one. In this situation, adaptation would virtually stop if $a(n)$ were  allowed to grow unchecked.
 To avoid this problem,  the auxiliary parameter $a(n)$ is  restricted by saturation to a symmetric interval $[-a^{+},\; a^{+}]$, which ensures
a minimum level of adaptation\footnote{A common choice in the literature is $a^{+}=4$.} \cite{Arenas06,AzpicuetaMLSP08}.
This truncation procedure constrains $\lambda(n)$ to the interval
 ${\rm sgm}[-a^{+}]\leq\lambda(n)\leq{\rm sgm}[a^{+}]$. In order to allow $\lambda(n)$ to take a value equal to zero or one, a scaled and shifted version of the sigmoid was proposed in \cite{LazaroTSP10},
  \begin{equation}
\lambda(n)=\varphi[a(n)] = \frac{{\rm sgm}[a(n)]-{\rm sgm}[-a^{+}]}{{\rm sgm}[a^{+}]-{\rm sgm}[-a^{+}]},\label{eq:activation2}
\end{equation}
which ensures that $\lambda(n)$ attains values $1$ and $0$ for $a(n)=a^{+}$ and $a(n)=-a^{+}$, respectively. Using \eqref{eq:activation2}, the {\em cvx-PN-LMS} rule becomes
\begin{equation}
a(n+1)= a(n)+\frac{\mu_a}{p(n)} e(n)[y_1(n)-y_2(n)]{\rm sgm}[a(n)]\left\{1-{\rm sgm}[a(n)]\right\}\Big\vert_{-a^+}^{a^+}.\label{eq:PNGDnewsgm}
\end{equation}
As we can see, $a(n)$ is truncated at every iteration to keep it inside the interval $[-a^{+},\; a^{+}]$. 

When deciding between affine or convex combinations with the above gradient-based rules, one should be aware of the following facts:
\begin{itemize}
\item Optimally adjusted affine combinations attain smaller EMSE than convex ones in some situations (cases 1 and 2 of Table \ref{tab:three_cases}).
\item Due to the constraints on the value of $\lambda(n)$, {\em cvx-LMS} and {\em cvx-PN-LMS} do not diverge, and large values of the step size $\mu_a$ can be used.
\item Gradient adaptation of the combination parameter implies that combinations introduce some additional gradient noise, which should be minimized with an adequate selection of $\mu_a$ or $\mu_\lambda$. In this sense, the factor $\lambda(n)[1-\lambda(n)]$ that appears in the adaptation of $a(n)$ with {\em cvx-PN-LMS} reduces the gradient noise when the mixing parameter becomes close to 0 or 1.
\end{itemize}
In our experience, the last two issues have an important impact on the performance of the combination, to the extent that convex combinations may actually be preferred over affine ones, unless in situations where significant EMSE gains can be anticipated for the affine combination from the theoretical analysis. The next subsection will compare the performance of these rules through simulations.

Before concluding the section, we should point out that other rules can be found in the literature for the adaptation of the mixing parameter, such as the {\em least squares} rule from \cite{AzpicuetaMLSP08}, the sign algorithm proposed in \cite{Shi14}, or a method relying on the ratio between the estimated errors of the filter components \cite{Bershad08}. Nevertheless, in the following we will restrict our discussion to the methods that have been presented in this section, which are the most frequently used in the literature.

\subsection{Convergence properties of combination filters}
\label{subsec:affine_vs_convex}

To examine the convergence properties of {\em aff-PN-LMS} and {\em cvx-PN-LMS}, we consider the combination of two normalized LMS (NLMS) filters with step sizes $\mu_1 = 0.5$ and $\mu_2 = 0.01$. The optimum solution is a stationary vector of length $7$, the covariance matrix of the input signal is $\Rbf = \frac{1}{7}\Ibf$, and the variance of the observation noise is adjusted to get an SNR of $20$ dB. Different step sizes have been explored for the combination: $\mu_a = [0.25, 0.5, 1]$ for {\em cvx-PN-LMS}, while $\mu_\lambda = \mu_a/800$ is used for {\em aff-PN-LMS} to get comparable steady-state error. Regarding these step-size values, we can see that the range of practical values for $\mu_a$ is
within the usual range of steps sizes used with normalized schemes, 
whereas for the affine combination much smaller values are required for comparable performance.  This fact simplifies the selection of the step size in the convex case.

\begin{figure}
	\centerline{{\begin{tabular}{c}\includegraphics[width=.4\columnwidth,trim= 0 0 0 0,clip=true,angle=0]{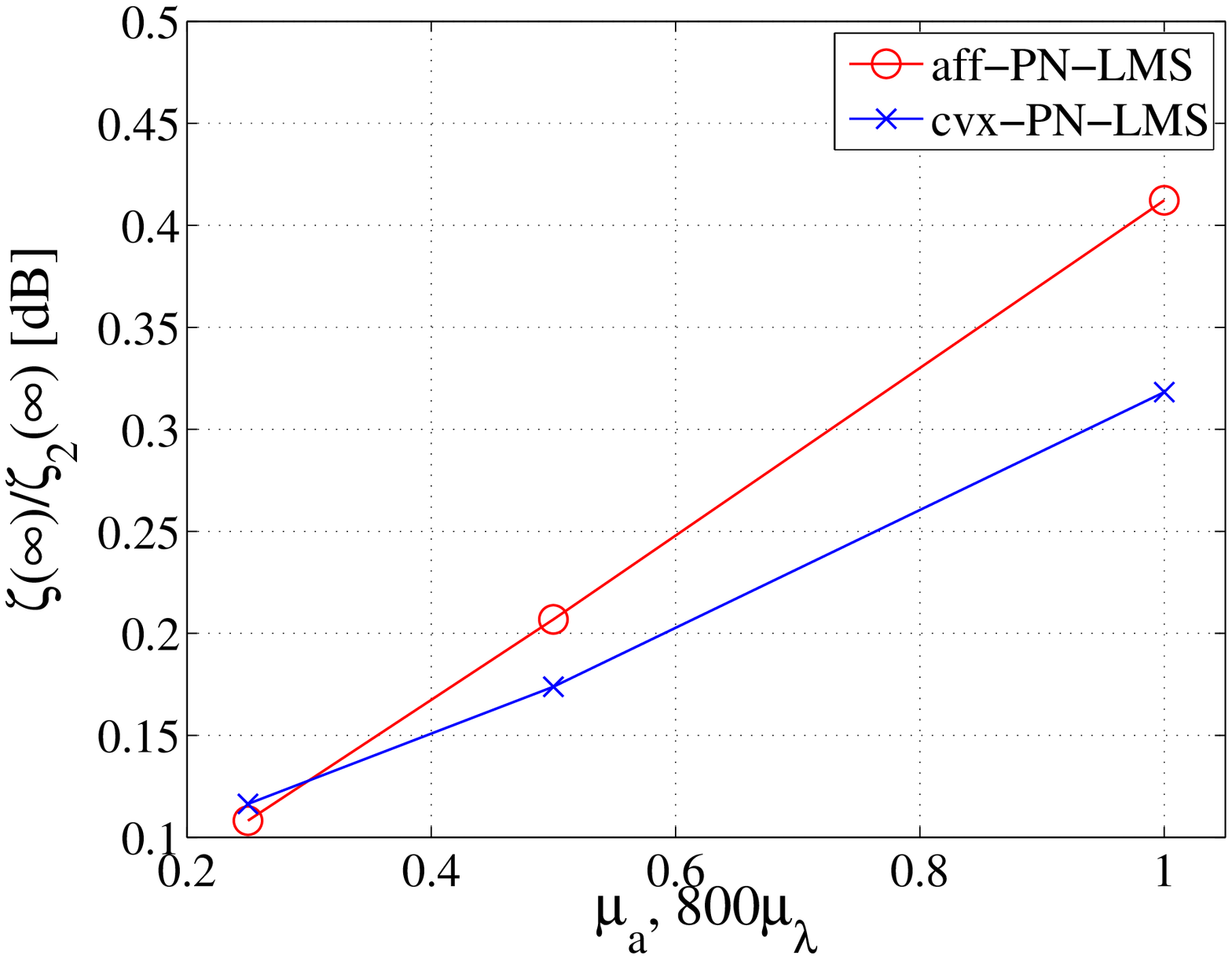}\\(a) \end{tabular}}} \centerline{{\begin{tabular}{c}\includegraphics[width=.4\columnwidth,trim= 0 0 0 0,clip=true,angle=0]{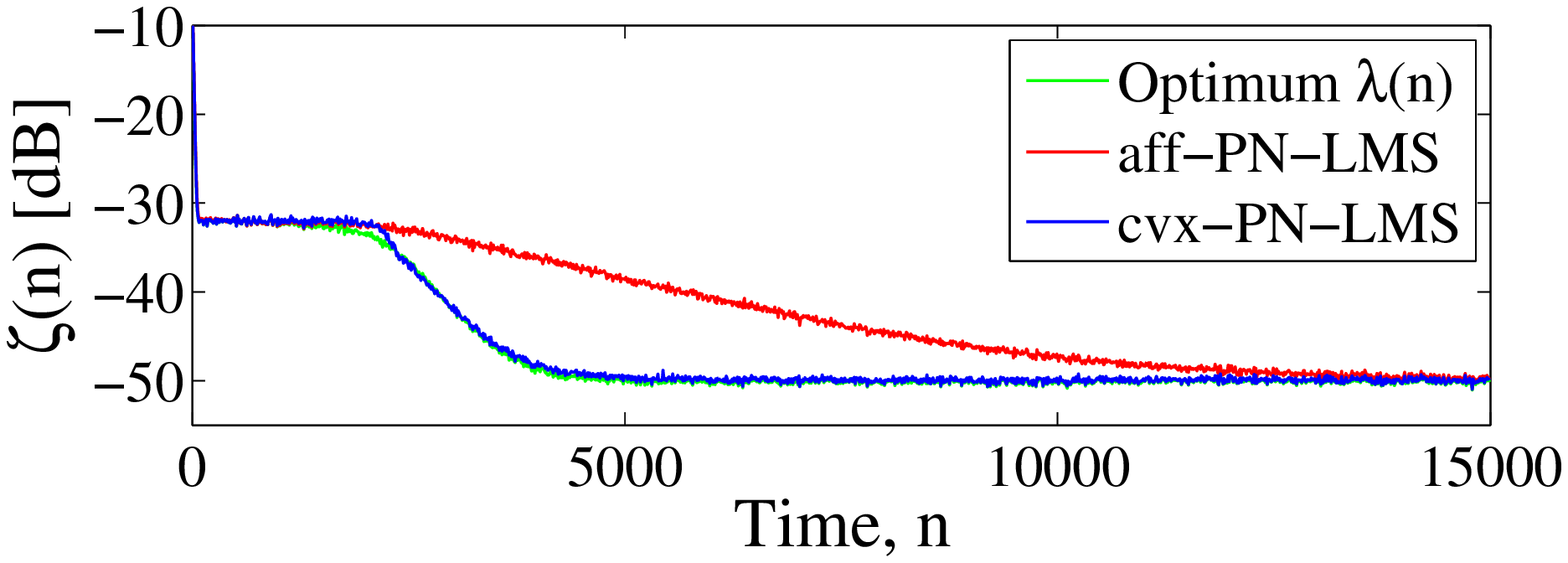}\\ (b) $\mu_a = 0.25$ \\ \includegraphics[width=.4\columnwidth,trim= 0 0 0 0,clip=true,angle=0]{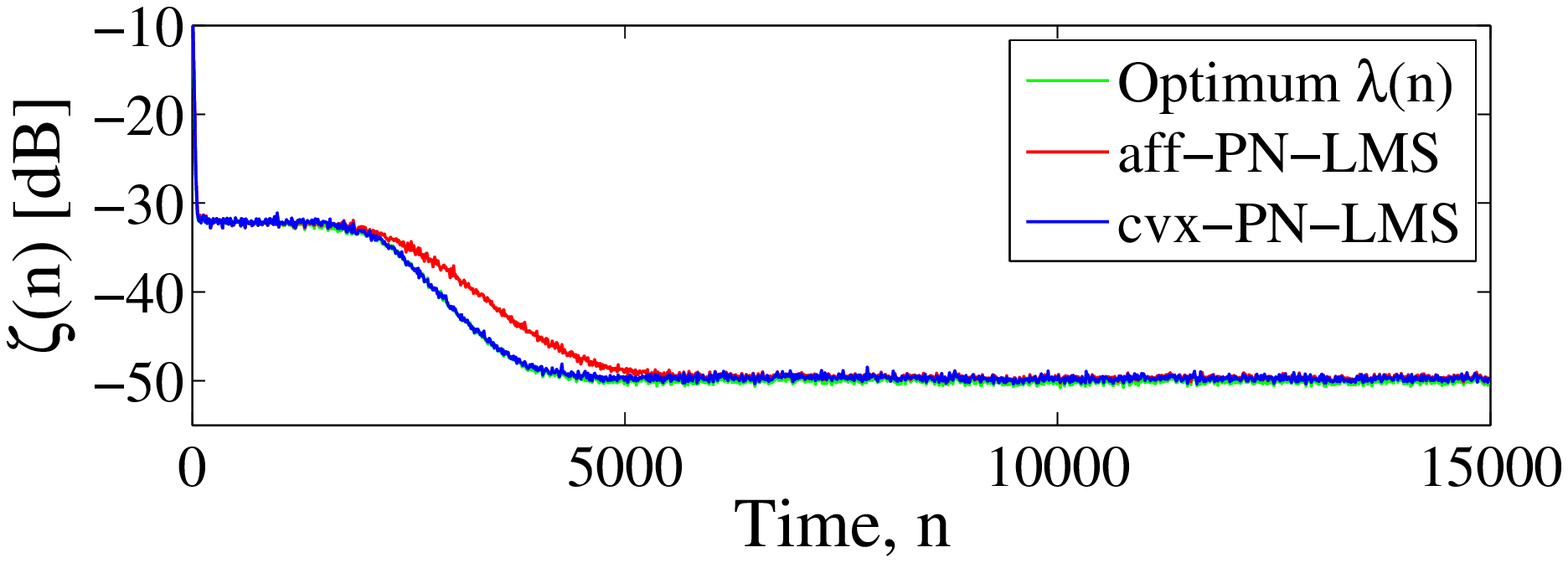} \\ (d) $\mu_a = 1$ \end{tabular}}{\begin{tabular}{c}\includegraphics[width=.4\columnwidth,trim= 0 0 0 0,clip=true,angle=0]{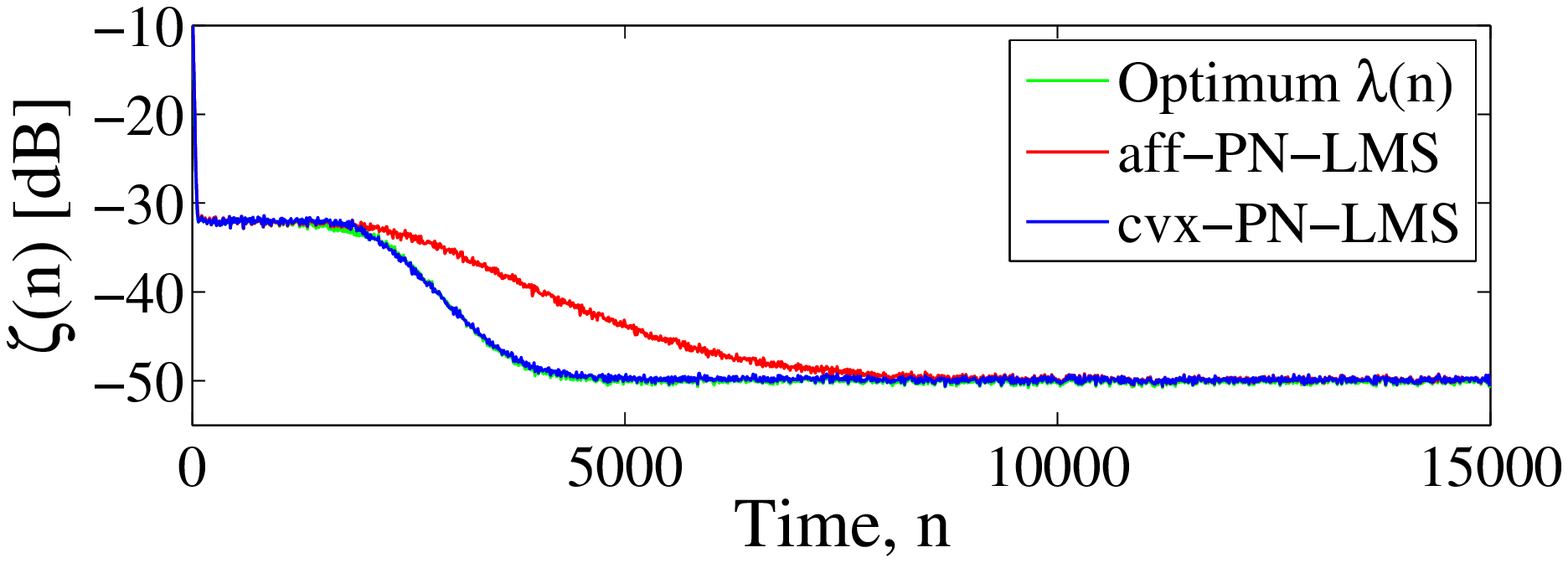}\\ (c) $\mu_a = 0.5$ \\ \includegraphics[width=.4\columnwidth,trim= 0 0 0 0,clip=true,angle=0]{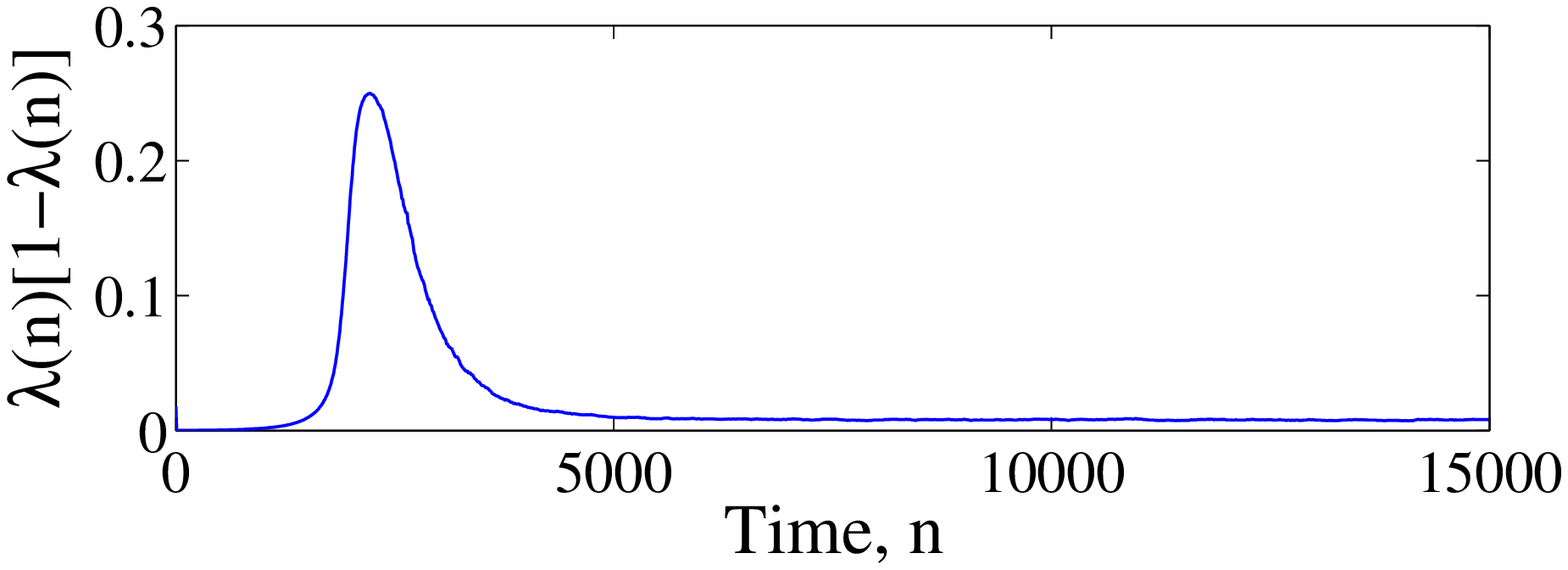} \\ (e) $\lambda(n) [1-\lambda(n)]$, $\mu_a = 0.5$ ({\em cvx-PN-LMS}) \end{tabular}}}
	\caption{Performance of {\em aff-PN-LMS} and {\em cvx-PN-LMS} when combining two NLMS adaptive filters with step sizes $\mu_1 = 0.5$ and $\mu_2 = 0.01$. (a) Steady-state excess error with respect to the slow filter component for three different values of the step sizes. Subfigures (b), (c), and (d) compare the convergence of the affine and convex rules (for three values of $\mu_a$ and $\mu_\lambda=\mu_a/800$) with the combination scheme using the optimum mixing parameter. (e) Evolution of factor $\lambda(n)[1-\lambda(n)]$ for {\em cvx-PN-LMS}. \label{fig:aff_cvx_conv}}
\end{figure}

Figure \ref{fig:aff_cvx_conv} illustrates the performance of affine and convex combinations averaged over $1000$ experiments. Subfigures (b)-(d) compare the convergence of both schemes with respect to the optimum selection of the mixing parameter given by \eqref{eq:lambda_opt}. In all cases, the combination schemes converge first to the EMSE level of the fast filter ($-30$ dB), and after a while follow the slow component to get a final EMSE of around $-50$ dB. It is interesting to see that {\em cvx-PN-LMS} shows near optimum selection of the mixing parameter for all three values of $\mu_a$, while the affine combination may incur a significant delay, especially for the smallest $\mu_\lambda$. Subfigure \ref{fig:aff_cvx_conv}-(a) plots the excess steady-state error of both schemes with respect to $\zeta_2(\infty)$ as a function of the step size, and shows that this faster convergence of {\em cvx-PN-LMS} with respect to {\em aff-PN-LMS} is not in exchange of larger residual error.

The fact that {\em cvx-PN-LMS} has the ability to switch rapidly between the fast and slow filter components while at the same time minimizing the residual error in steady state is due to the incorporation of the activation function, whose derivative propagates to the update rule. In other words, we can view the effective step size of {\em cvx-PN-LMS} as being $\mu_a \lambda(n) [1-\lambda(n)]$ (see Table \ref{tab:lblearning}), and the evolution of the multiplicative factor, represented in Figure \ref{fig:aff_cvx_conv}-(e), shows that this effective step size becomes large when the combination needs to switch between filter components, while becoming small in steady state, thus minimizing the residual error after convergence is complete. 

\subsection{Benefits of power normalized updating rules}
\label{subsec:beneritsPN}
In order to illustrate the benefits of power normalized updating rules, we compare the behavior of a convex combination of two NLMS filters with step sizes $\mu_1 = 0.5$ and $\mu_2 = 0.01$, employing both the {\em cvx-LMS} rule and its power normalized version ({\em cvx-PN-LMS}) for the combination layer. The optimum solution is a length-30 nonstationary vector, which varies according to the random-walk model given by~\eqref{eq:randomwalk}.
The covariance matrices of the change of the optimum solution and of the input signal are
 given respectively by  $\mathbf{Q}=\sigma_q^2\mathbf{I}$ and  $\mathbf{R} = \frac{1}{30}\mathbf{I}$. The variance of the observation noise
is adjusted to get different SNR levels.
The step size for the \emph{cxv-LMS} rule has been set to $\mu_a= 1000$ and for its power normalized version (\emph{cvx-PN-LMS}), we have set $\mu_a=1$.

\begin{figure}[t]
 	\centering
 	\includegraphics[width=1\columnwidth,angle=0,clip=true,trim=1cm 14.1cm 2cm 7.25cm]{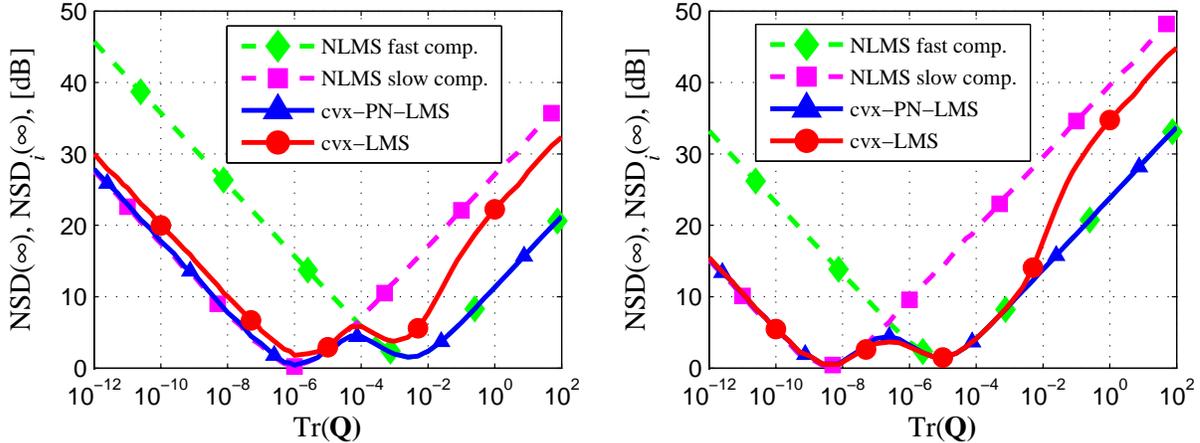}
 	\caption{Steady-state NSD of a convex combination of two NLMS filters with different step sizes as a function of the speed of changes
 		of the optimum solution. The figures illustrate the NSD incurred by the individual NLMS filters and by their convex combinations.
 		NLMS step sizes are $\mu_1=0.5$ and $\mu_2=0.01$; combination rules: \emph{cvx-LMS} ($\mu_a=1000$) and \emph{cvx-PN-LMS} ($\mu_a=1$,\;$\eta=0.9$); signal-to-noise ratios:
 		(left) $\rm{SNR}=5~{\rm dB}$ and (right) $\rm{SNR}=30~{\rm dB}$. All results were averaged over 50000 iterations once the algorithms
 		reached steady state and over 50 independent runs.
 		\label{fig:PNtracking}}
 \end{figure}

Figure~\ref{fig:PNtracking} shows the steady-state NSD of the individual filters and of their convex combinations obtained with
the \emph{cxv-LMS} rule and with the \emph{cvx-PN-LMS} scheme, as a function of the speed of changes
of the optimum solution [${\rm Tr}(\mathbf{Q})=30\sigma_q^2$].
The left panel considers an SNR of 5~dB and the right panel, an SNR of 30~dB.
We can observe that the combination scheme
with  the \emph{cxv-LMS} rule results in a
suboptimal performance when the optimum solution changes very fast
[for large ${\rm Tr}(\mathbf{Q})$]. This is due to the fact that both component
filters are incurring a very significant error, resulting
in a non-negligible gradient noise when updating the auxiliary parameter $a(n)$ with \emph{cxv-LMS} \cite{AzpicuetaICASSP08}. 
The performance of \emph{cxv-LMS} also
degrades, whatever the value of  ${\rm Tr}(\mathbf{Q})$, as the SNR decreases.
On the other hand, when the \emph{cvx-PN-LMS} rule
 is employed, the combination shows a very stable operation,
and behaves as well as the best component filter not
only for any SNR, but also for all values of ${\rm Tr}(\mathbf{Q})$. 
A similar behavior is also observed when we compare the \emph{aff-LMS}  rule  to its power normalized version
(\emph{aff-PN-LMS}) to update $\lambda(n)$ in affine combinations  \cite{Candido10}.

\subsection{Combination schemes versus VSS LMS}
\label{subsec:vss}
We have illustrated so far how combination schemes can be exploited to provide filters with improved performance, relying on adaptive rules to learn the most adequate value for the mixing parameter. A reasonable question is whether the same improvement could be achieved by an individual but more complex filter structure. Although the answer to this question is generally positive, these more complex solutions normally require further statistical knowledge about the filtering scenario that is normally unavailable, while combination approaches offer a more versatile and robust approach. To illustrate this point, we will examine again the identification of a time-varying solution. 

Adaptive filters usually have two conflicting goals: on one hand, to track variations of the parameter vector they are trying to estimate and, on the other hand, to suppress measurement noise and errors due to undermodeling.
The first goal requires a fast filter, one that quickly corrects any mismatch between the input and the estimates, while the second goal would benefit from a slower filter that averages out measurement noise. We can emphasize one goal over the other by the choice of a design variable such as the step size in LMS or the forgetting factor in RLS. However, in a nonstationary environment, the optimum choice of the step size or the forgetting factor will continuously change, which has motivated the proposal of different methods to update the step size (or forgetting factor) ---see e.g., \cite{Harris86,Kwong92,Mathews93,Aboulnasr97,Costa06}.
Most existing variable step size (VSS) algorithms implement procedures that rely on the available input and error signals. However, in the absence of additional information (e.g., background noise level, whether the filter is still converging or has reached steady state), these algorithms can fail to identify the most convenient step size, especially if the filtering conditions change over time. 

\begin{figure*}
\centerline{\includegraphics[width=0.8\columnwidth,trim= 4cm 11cm 4cm 11cm,clip=true]{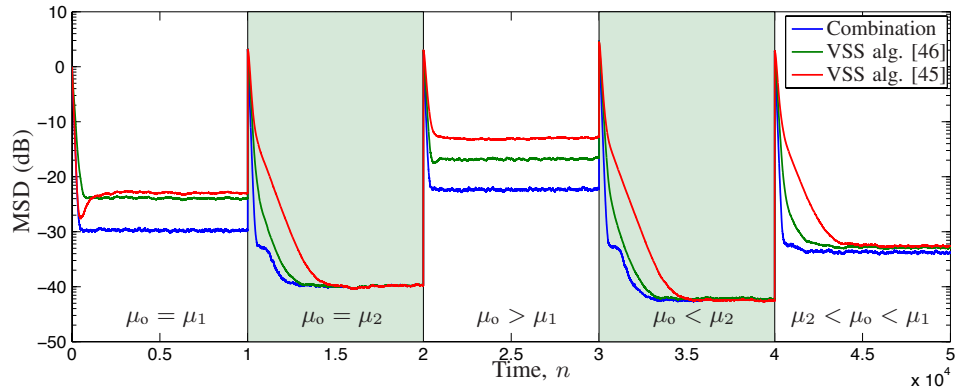}}
\caption{Mean-square deviation (MSD) obtained by VSS algorithms and by the convex combination of two LMS filters with $\mu_1 = 0.01$ and $\mu_2 = 0.001$. The value of $\sigma_q^2$ changes with time, and the corresponding optimum LMS step sizes ($\mu_\text{o}$) are indicated in the figure.\label{stepsize_fig}}
\end{figure*}

As an alternative to the VSS algorithms, combinations of adaptive filters with different step sizes (or forgetting factors) can be particularly useful. One advantage of this approach is that it performs reasonably well for large variations in input signal-to-noise ratio and speed of change of the true parameter vector. Figure \ref{stepsize_fig} compares the performance, in an identification problem, of the robust variable step-size methods proposed in \cite{Aboulnasr97} and \cite{Costa06} against the performance of convex combination of two LMS adaptive filters with different step sizes.  The fast filter uses $\mu_1=0.01$ and the slow filter, $\mu_2=0.001$.    
The figure shows the mean-square deviation (MSD) of a 10-tap filter in a tracking situation like the one described by \eqref{eq:randomwalk}, with ${\bf Q} = \sigma_q^2 {\bf I}$, for time-varying $\sigma_q^2$ as indicated in the figure.   In order to provide a fair comparison, the parameters for the VSS methods were optimized for step sizes in the range $[\mu_2, \mu_1]$, and the figure shows the performance of the three methods for five different values of the optimum step size $\mu_\text{o}$ for a single LMS filter: from less than $\mu_2$ to more than $\mu_1$.
It is seen that the convex combination scheme is able to deliver the best possible performance in all conditions. In other words, it shows a more robust performance given the lack of knowledge about the true value of $\sigma_q^2$, and attains an overall performance that could not be achieved by a standard LMS filter or by the VSS schemes.



\section{Communication between component filters}\label{sec:transfer}
One simple way to improve the performance of combination schemes is to allow some communication between the component filters.  Since the component filters are usually designed to perform better in different environmental conditions, information from the best filter in a given set of conditions can be used to boost the performance of the other filters.  In some situations, the gain for the overall output can be significant.

\begin{figure}[t]
\centerline{\includegraphics[width=.45\columnwidth,trim=6cm 6.5cm 6cm 4cm,clip=true] {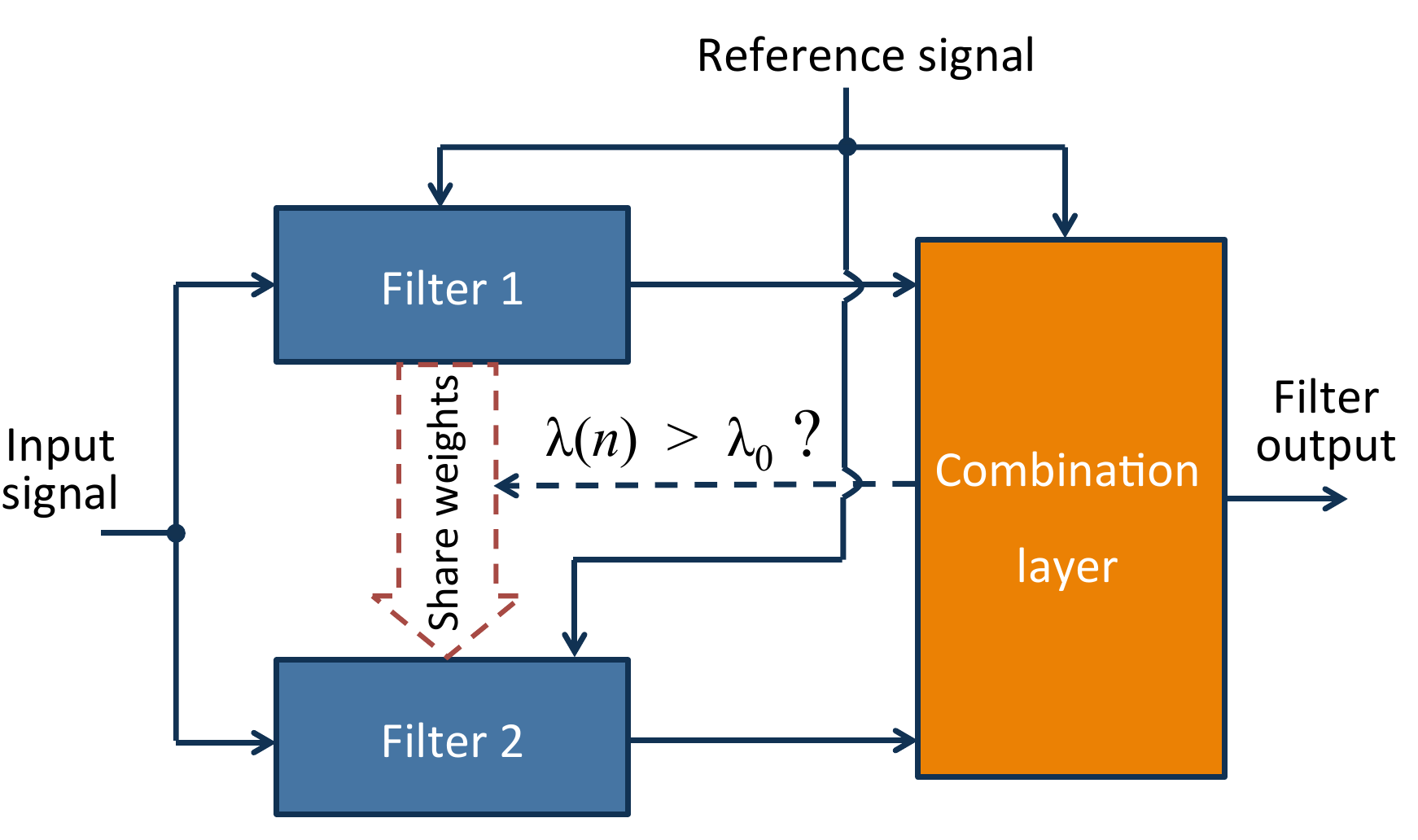}\ \ \includegraphics[width=.45\columnwidth,trim=6cm 6.5cm 6cm 4cm,clip=true]{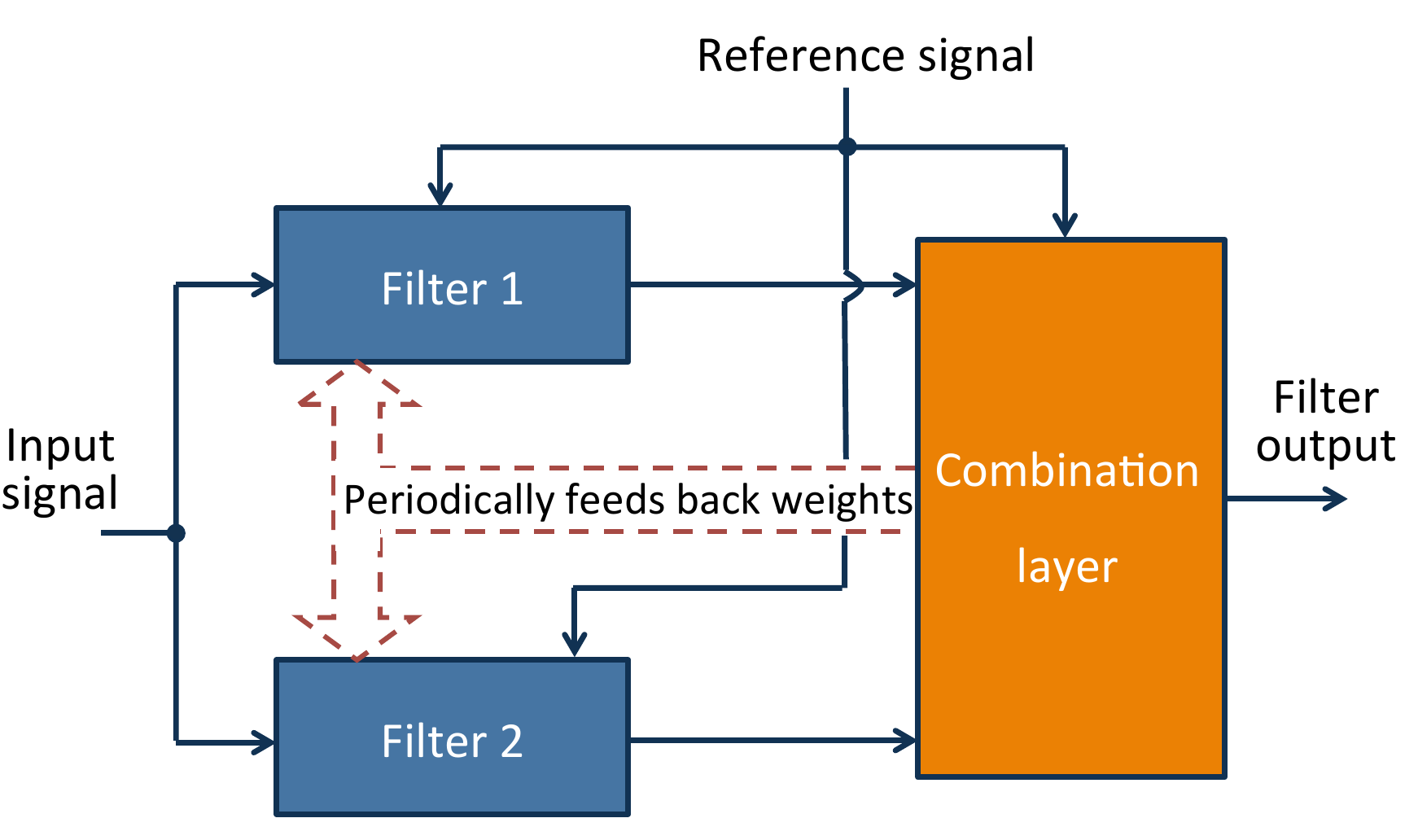}}
\caption{Communication between sub-filters.  Left panel: weight transfer; right panel: weight feedback.\label{fig:combtransf}}
\end{figure}

We describe here two approaches for inter-filter communication: the weight transfer scheme of \cite{arenas-garciaElsevier06,Nascimento12}, and the weight feedback of \cite{Chamon2012}.  Both are applicable when the component filters have the same length.  The difference between these methods is that the first allows only communication from one of the filters to the other, whereas the second approach feeds back the overall combined weights to the component filters, as depicted in Figure~\ref{fig:combtransf}.

\begin{figure}[t]
\centering\includegraphics[width=0.8\columnwidth]{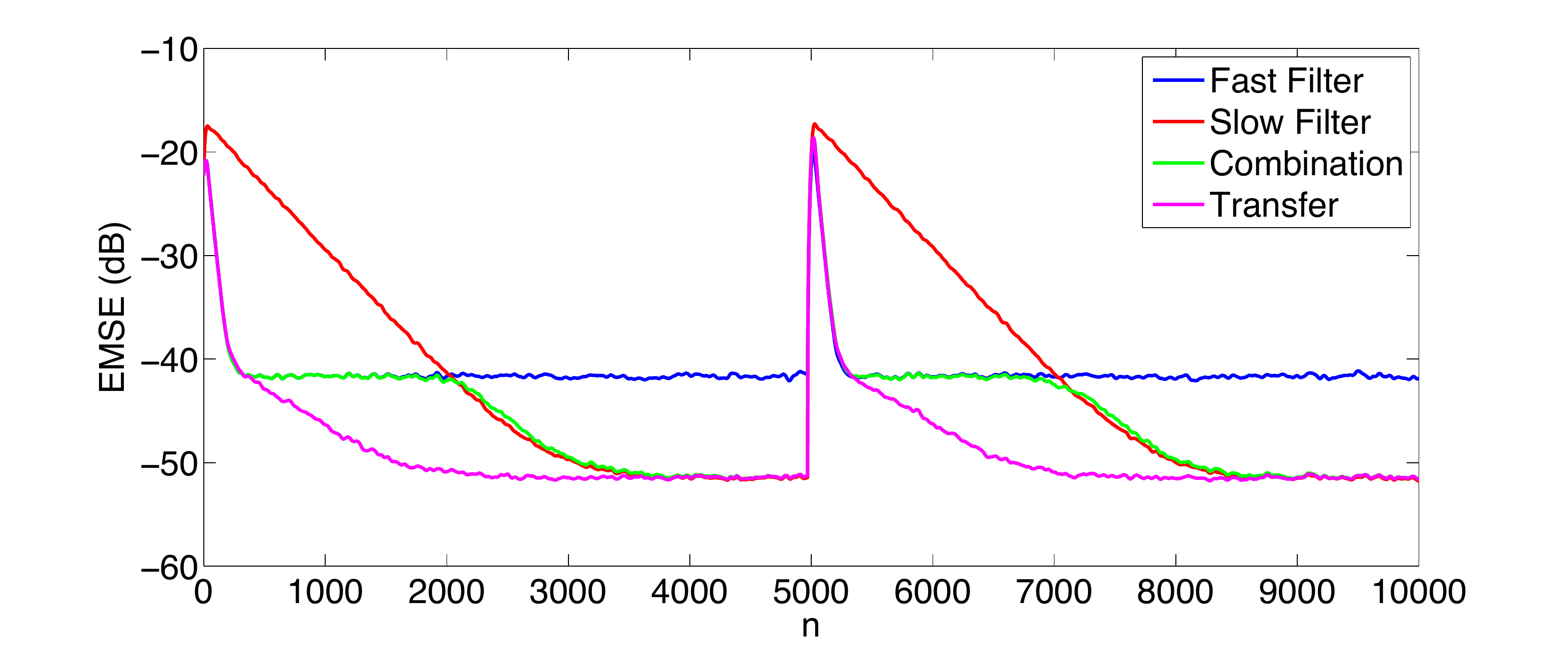}
\caption{Performance of weight transfer schemes.  The plots show the EMSE obtained with the component filters and standard convex combination without weight transfer, and the second method of Table~\ref{tab:weight.transfer}, using $N_0=2$ and $\lambda_0=0.982$. The curves are averages of $L=1500$ simulations.  The components are two NLMS algorithms, one using $\mu_1=0.1$, the other using $\mu_2=0.01$.\vspace{-.3cm}}\label{fig:transfer.example2}
\end{figure}

Consider for example the combination of two component filters, one designed for a fast-changing environment (``fast'' filter), and the other for a slow-changing environment (``slow'' filter).  In this situation, during the initial convergence, or after an abrupt change in the optimum weight vector $\wo(n)$, the slow filter may lag behind the fast filter (see Figure~\ref{fig:transfer.example2}).  After the fast filter reaches the steady state, the combination would need to wait until the slow filter catches up.  This can be avoided by either leaking \cite{arenas-garciaElsevier06} or copying \cite{Nascimento12} the fast filter coefficients to the slow filter at the appropriate times.

The good news is that the mixing parameter is already available to help decide when to transfer the coefficients ---we only need to check if $\lambda(n)$ is close to one (so that the overall output is essentially the fast filter).  Therefore, the idea is to choose a threshold $\lambda_0$, and modify the slow filter update whenever $\lambda(n)\ge \lambda_0$, as shown in Table~\ref{tab:weight.transfer}. The two methods are similar: in the gradual transfer scheme of \cite{arenas-garciaElsevier06}, whenever $\lambda(n)$ is larger than the threshold, we allow the fast filter coefficients to gradually ``leak'' into the slow filter.   The other method simply copies the fast filter coefficients to the slow filter, and thus has a smaller computational cost; however, the condition for transfer has to be modified.  If we allowed the fast filter coefficients to be copied to the slow filter whenever $\lambda(n)\ge \lambda_0$, the combination would be stuck forever at the fast filter (the slow filter adaptation would be irrelevant). For this reason the transfer is only allowed at periodic intervals of length $N_0\ge 2$.
\begin{table}
\caption{Inter-filter communication schemes (exemplified for combinations of two filters).  }\label{tab:weight.transfer}
\begin{center}
\begin{tabular}{llll}
\toprule
& Gradual transfer \cite{arenas-garciaElsevier06} & Simple copy \cite{Nascimento12} & Feedback \cite{Chamon2012}\\
\midrule
Filter communication & $\w_2(n) \leftarrow \ell\w_2(n)+ (1-\ell)\w_1(n)\;\;\;$ & $\w_2(n) \leftarrow \w_1(n)$ & $\w_1(n), \w_2(n) \leftarrow \w(n)$ \\
Condition & $\lambda(n) \ge \lambda_0$ & $\lambda(n)\ge \lambda_0$ and $n = k N_0\;\;\;\;\;$ & $n=k N_0$ \\
Parameters & $0<\ell<1; ~~0\ll \lambda_0<1$ & $N_0 \ge 2; ~~0\ll \lambda_0<1$ & $N_0\ge 2$ \\
\bottomrule
\end{tabular}
\end{center}
\end{table}

The weight feedback scheme, on the other hand, copies the combined weight vector $\w(n)=\lambda(n)\w_1(n)+(1-\lambda)\w_2(n)$ periodically to $\w_1(n)$ and $\w_2(n)$, as shown in Figure~\ref{fig:combtransf} and Table~\ref{tab:weight.transfer}.  This method is simpler to set up, since only one parameter needs to be tuned, but requires the explicit computation of $\w(n)$ at intervals of $N_0$ samples.  Since $N_0$ can be chosen reasonably large, the additional cost per sample is modest. Even though these methods introduce additional parameters that need to be fixed, their selection is not very problematic and the transfer methods show a behavior rather robust with respect to them (a detailed discussion is not included here, but the interested readers are referred to \cite{arenas-garciaElsevier06,Nascimento12,Chamon2012}).

Figure~\ref{fig:transfer.example2} provides an example of the operation of the simple-copy method for length $M=7$ filters (the other methods have similar performance).  The value of $\wo(n)$ changes abruptly at the middle of the simulation ($n=5\times 10^4$).  

\section{Combination of several adaptive filters}
Up to now, we have considered combinations of just two adaptive filters. However, combining $K$ adaptive filters (with $K>2$) makes it possible to further increase the robustness and versatility of combination schemes. The combination of an arbitrary number of filters can be mainly used to:

\begin{itemize}
\item{Simplify the selection of a parameter. For instance, paying attention to the tracking scenario depicted in the left plot of Fig. \ref{LMS_LMS_theo}, the robustness of the scheme would be increased for $\tr\{\Qbf\} >  10^{-3}$ if a third filter with step size equal to one is incorporated to the combination \cite{AzpicuetaISCAS10}.}
\item{Alleviate several compromises simultaneously. For instance, regarding the selection of the step size, $\mu$, and the length of an adaptive filter, $M$, we can combine four adaptive filters with settings $\{\mu_1, M_1\}$, $\{\mu_2, M_1\}$, $\{\mu_1, M_2\}$ and $\{\mu_2, M_2\}$. Another example was proposed in \cite{Azpicueta11} and it is included in Section XI, where a combination of several filters (linear and nonlinear) is designed to alleviate simultaneously the compromise related with the step size selection, and with the presence or absence of nonlinearities in the filtering scenario.}
\end{itemize}
In the literature, two different approaches for the combination of several adaptive filters have been proposed, both for affine and convex approaches. These schemes differ in the topology employed to perform the combination as described below.


\subsection*{The hierarchical scheme}

\begin{figure}[t]
\centerline{\includegraphics[width=.55\columnwidth,trim= 4cm 3cm 4cm 3.5cm,clip=true]{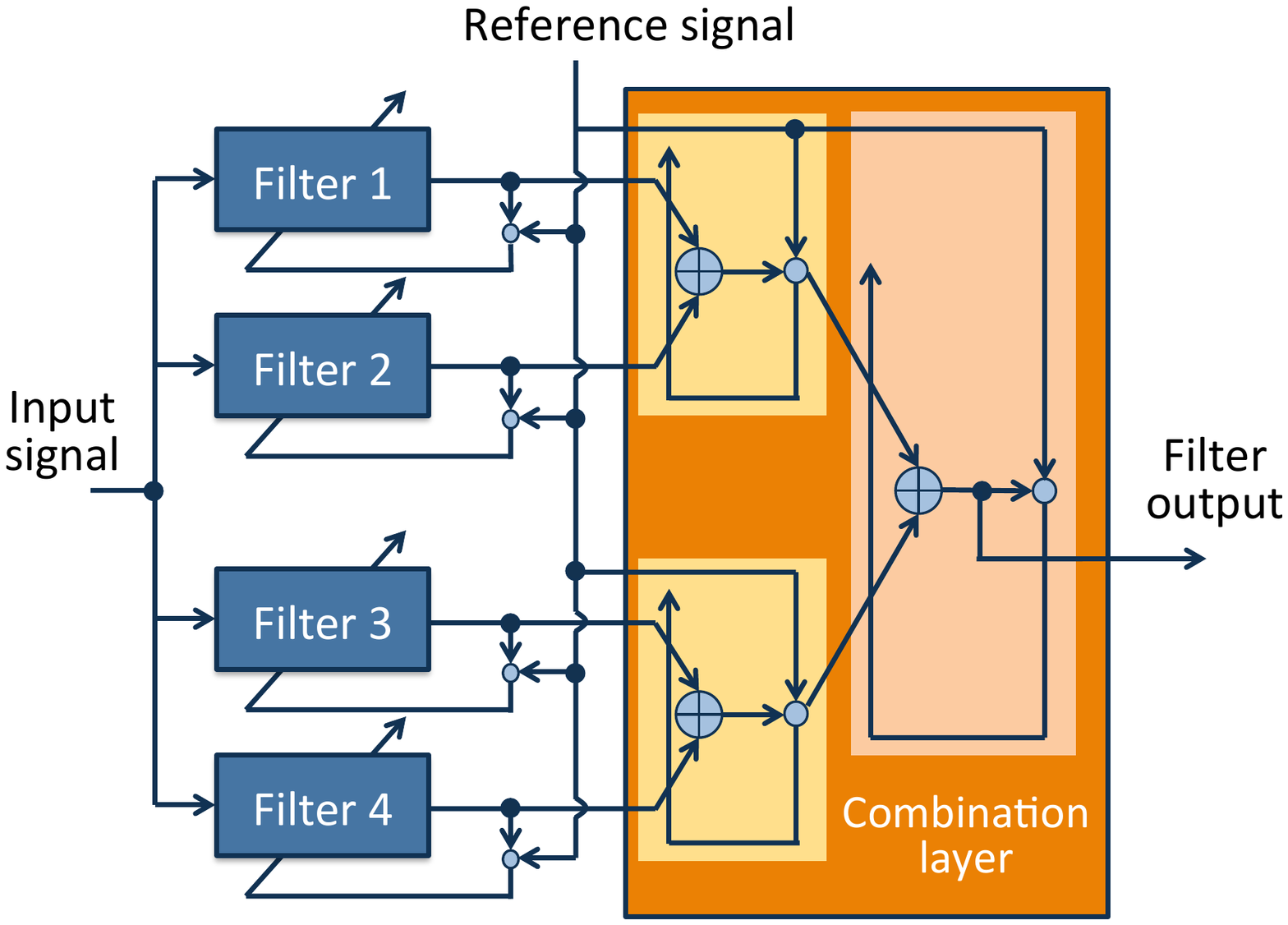}\includegraphics[width=.55\columnwidth,trim= 4cm 3cm 4cm 3.5cm,clip=true]{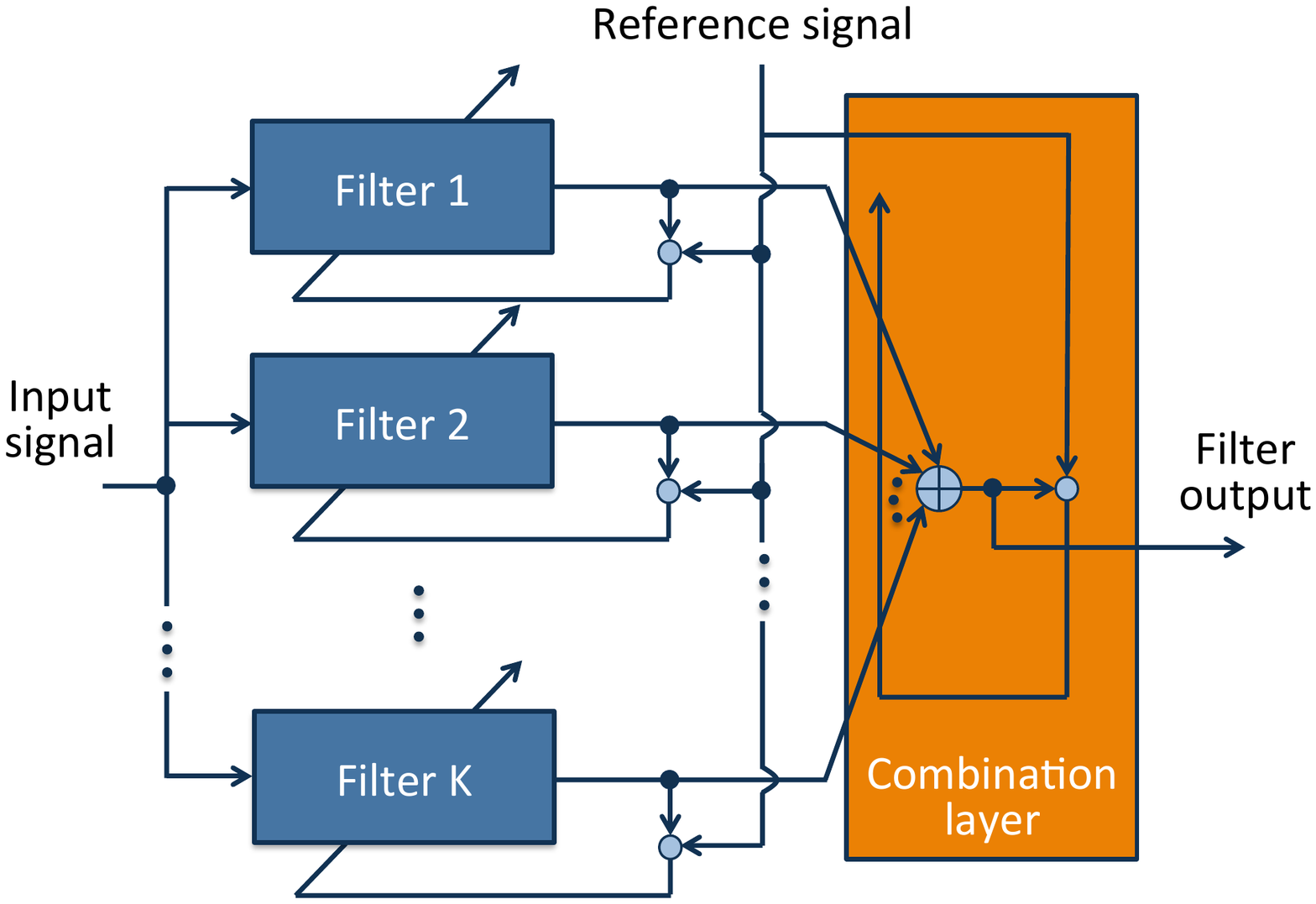}}
{\hspace{2cm} (a) {\em Hierarchical} scheme \hspace{5.5cm}(b) {\em One-layer} scheme}
\caption{Hierarchical and one-layer combination schemes for an arbitrary number of filters $K$. (a) Hierarchical scheme: In the first layer, we carry out two combinations of two adaptive filters with complementary settings, and their outputs are combined in the upper layer. (b) Combination of $K$ adaptive filters following the one-layer approach. The affine schemes of \cite{AzpicuetaISCAS10} and \cite{Kozat10} adapt $K-1$ mixing parameters, whereas the convex combination of \cite{arenas-garciaInstru05,Kozat10} requires the adaptation of $K$ parameters.}
\label{fig:multifilter}
\end{figure}

This approach combines $K$ adaptive filters employing different layers, where only combinations of two elements are considered at a time. For instance, the output of a hierarchical combination of four filters depicted in Fig. \ref{fig:multifilter}(a) reads:
\begin{equation}
y(n) = \lambda_{21}(n)\{\lambda_{11}(n)y_1(n) + [1- \lambda_{11}(n)] y_2(n)\} +  [1-\lambda_{21}(n)]\{\lambda_{12}(n)y_3(n) + [1-\lambda_{12}(n)] y_4(n)\},
\label{eq_hier}
\end{equation}
where $\lambda_{ij}(n)$ refers to the mixing parameter of the $j$-th combination in the $i$-th layer. All mixing parameters are adapted in order to minimize the power of the local combined error. For their update, we can follow the same adaptive rules (convex or affine) reviewed previously for the case of a combination of two filters.

\subsection*{The one-layer scheme}


We can combine an arbitrary number of filters employing an alternative approach based on one-layer combination as depicted in Fig. \ref{fig:multifilter}(b). Focusing on affine combinations of $M$ adaptive filters with outputs $y_k(n)$, with $k=1,...,K$, references \cite{Kozat10} and \cite{AzpicuetaISCAS10} proposed two one-layer combination schemes whose output is given by
\begin{equation}
y(n) = \sum_{k=1}^{K-1}{\lambda_k(n)y_k(n)} + \left[1- \sum_{k=1}^{K-1}\lambda_k(n)\right]y_K(n).
\label{eq_LS}
\end{equation}
Different adaptive rules were proposed in the literature to update $\lambda_k(n)$, with $k=1,...,K-1$, following, for instance, LMS or RLS approaches \cite{Kozat10}, or estimating the $K-1$ affine mixing parameters as the solution of a least-squares problem \cite{AzpicuetaISCAS10}. 
The incorporation of convex combination constraints forces the inclusion of an additional mechanism (similar to the sigmoidal activation) to make all mixing parameters remain positive and add up to one. This scheme was proposed by \cite{arenas-garciaInstru05} and \cite{Kozat10} as an extension to the standard convex combination of two adaptive filters, and obtains the combined output as  
\begin{equation}
y(n)= \sum_{k=1}^{K}\lambda_k(n)y_k(n).
\label{eq_outsoft}
\end{equation}
As in the case of combining two filters, instead of adapting directly the $K$ mixing parameters, $K$ auxiliary parameters $a_k(n)$ are updated following a gradient descent algorithm. The relation between $\lambda_k(n)$ and $a_k(n)$ is based on the {\em softmax} activation function
\begin{equation}
\lambda_k(n) = \frac{{\text{exp}}[a_k(n)]}{\sum_{j=1}^K {\text{exp}}[a_j(n)]},\;\;\;\text{for}\; k=1,...,K.
\label{softmax}
\end{equation}
This activation function is a natural extension of the {\em sigmoid} used in the binary case to map several real parameters to a probability distribution \cite{Bishop95}, as required by a convex combination, where all parameters must remain positive and sum up to one.


Although both multifilter structures can improve the performance beyond the combination of just two filters, one useful characteristic of hierarchical schemes is its ability to extract more information about the filtering scenario from the evolution of the mixing parameters, since each combination usually combines two adaptive filters that only differ in the value of a setting (step size, length, etc.), and the combination parameter selects the best of both competing models.


\section{Reduced cost combinations}
Combination schemes require running two or more filters in parallel, which may be a concern in applications in which computational cost is at a premium. In many situations, however, the additional computational cost of adding one or more filters can be made just slightly higher than the cost of running a single filter.  In this section we describe a few methods to reduce the cost of combination schemes.

\subsection*{Use a low-cost filter as companion to a high-cost one}

Although straightforward, several useful results fall in this class.  
 For example, consider the previously mentioned case of a combination of an RLS and an LMS filter. This structure enhances the tracking performance of a single RLS filter, and only requires a modest increase in computational complexity. A lattice implementation of RLS has a computational cost of about $16M$ multiplications, $8M$ additions, and $8M$ divisions \cite{Sayed08}.  Compared with the approximately $2M$ additions and $2M$ multiplications  required by LMS, the combination scheme will increase the cost by only about 10\% over that of a single RLS. For other stable implementations of RLS, such as QR methods, the computational cost is even higher, so the burden of adding an LMS component would be almost negligible\footnote{Note that, for their operation, combination schemes require just the outputs of component filters (see Section \ref{sec:lambda_learning}), so lattice or QR implementations of RLS can readily be incorporated into combination schemes without modifications.}.

Other situations in which it is useful to combine a low-cost and a higher-cost filter are:  the combination of a simple linear filter with a Volterra nonlinear filter, as described in Section~\ref{sec:acoustic.echo}; and the use of a short filter combined with a longer filter, as described for example in \cite{Azpicueta-Ruiz2014}.  The short filter is responsible for tracking fast variations in the weight vector, whereas the long filter is responsible for guaranteeing a small bias in times of slower variation of the optimum weight vector.

\subsection*{Parallelization}
Since the component filters are running independently (apart from occasional information exchanges as in Section~\ref{sec:transfer}), combination schemes are easy to implement in parallel form.
In an FPGA implementation, for example, this means that implementing a combination of filters will not require a faster clock rate, compared to running a single filter.  Depending on the hardware in which the filters are to be implemented, this is an important advantage.

\subsection*{Take advantage of redundancies in the component filters}
  This approach can provide substantial gains in some situations, but depends on the kind of filters that are being combined. For example, transform-domain or subband filters can share the transform or filter bank blocks that process the input signal $\uu(n)$.

A second example is based on the following observation \cite{Nascimento2013a}: if filters using the same input regressor $\uu(n)$ are combined, the weight estimates $\w_i(n)$ will be, most of the time, similar.  Therefore, one could update only one of the filters (say, $\w_1(n)$), and the \emph{difference} between this filter and the others.
The important observation is that, if the entries of $\w_1(n)$ and $\w_2(n)$ are real values ranging from (say) $-1$ to $+1$, the range of the entries of $\boldsymbol{\delta}\w_2(n)=\w_2(n)-\w_1(n)$ will most of the time be smaller, from $-2^{-B_{\text{c}}}$ to $+2^{-B_{\text{c}}}$, for some number of bits $B_{\text{c}}>0$.  If the filters are implemented in custom or semi-custom hardware, this observation can be used to reduce the number of bits allocated to $\boldsymbol{\delta}\w_2(n)$, reducing the complexity of all multiplications related to the second filter, both for computing the output and for updating the filter coefficients. In summary, this idea amounts to making the filters share the most significant bits in all coefficients, and adapting them with the step size of the fast filter. During convergence, when the difference between the filters is expected to be more significant, disregarding the most significant bits of the difference filter has the effect of working just as a weight transfer mechanism from the fast to the slow filter.

In order to take full advantage of the reduced wordlength used to store $\boldsymbol{\delta}\w_2(n)$, the output of the second filter and the difference filter update should be computed as shown in Table~\ref{tab:reduced.cost}. Algorithm's steps 3 and 6 can be computed with a reduced wordlength.  As can be seen, all the costly operations related to the second filter are computed with a reduced wordlength.  Using this technique, it is possible to use for the difference filter a wordlength of about half that of the first filter, with little to no degradation in performance, compared to using full wordlength for all variables.
\begin{table}
\caption{Reduced-cost combination through difference filter (using two LMS filters as example)}\label{tab:reduced.cost}
\centering
\begin{tabular}{lll}
	\toprule
	Algorithm Step & Equation & Wordlength \\
	\midrule
	1 --- First filter error: & $e_1(n)=d(n)-\uu^{\top}(n)\w_1(n)$ & Full \\
	2 --- Update first filter: &
	$\w_1(n+1)=\w_1(n)+\mu_1 e_1(n)\uu(n)$ & Full \\
	3 --- Output of difference filter: & $\delta y_2(n)=\uu^{\top}(n)\boldsymbol{\delta}\w_2(n)$  & Reduced\\
	4 --- Second filter error: & $e_2(n)=e_1(n)+\delta y_2(n)$ & Full\\
	5 --- Output error: & $\lambda(n) e_1(n)+(1-\lambda(n))e_2(n)$ & Full\\
	6 --- Update difference filter: & $\e_{\delta}(n)=\mu_2 e_2(n)-\mu_1 e_1(n)$ & Reduced \\
	& $\boldsymbol{\delta}\w_2(n+1)=\boldsymbol{\delta}\w_2(n)+e_{\delta}(n)\uu(n)$ & \\
	\bottomrule
	\end{tabular}

\end{table}
%

\section{Signal modality characterization}
As we have already discussed, one of the main advantages of the combination approach to adaptive filters is its versatility. In the rest of this tutorial we illustrate the applicability of this strategy in different fields where adaptive filters are generally used.

\par
\begin{figure}
\centerline{\includegraphics[width=0.6\columnwidth,trim= 0cm 1cm 0cm 0cm,clip=true]{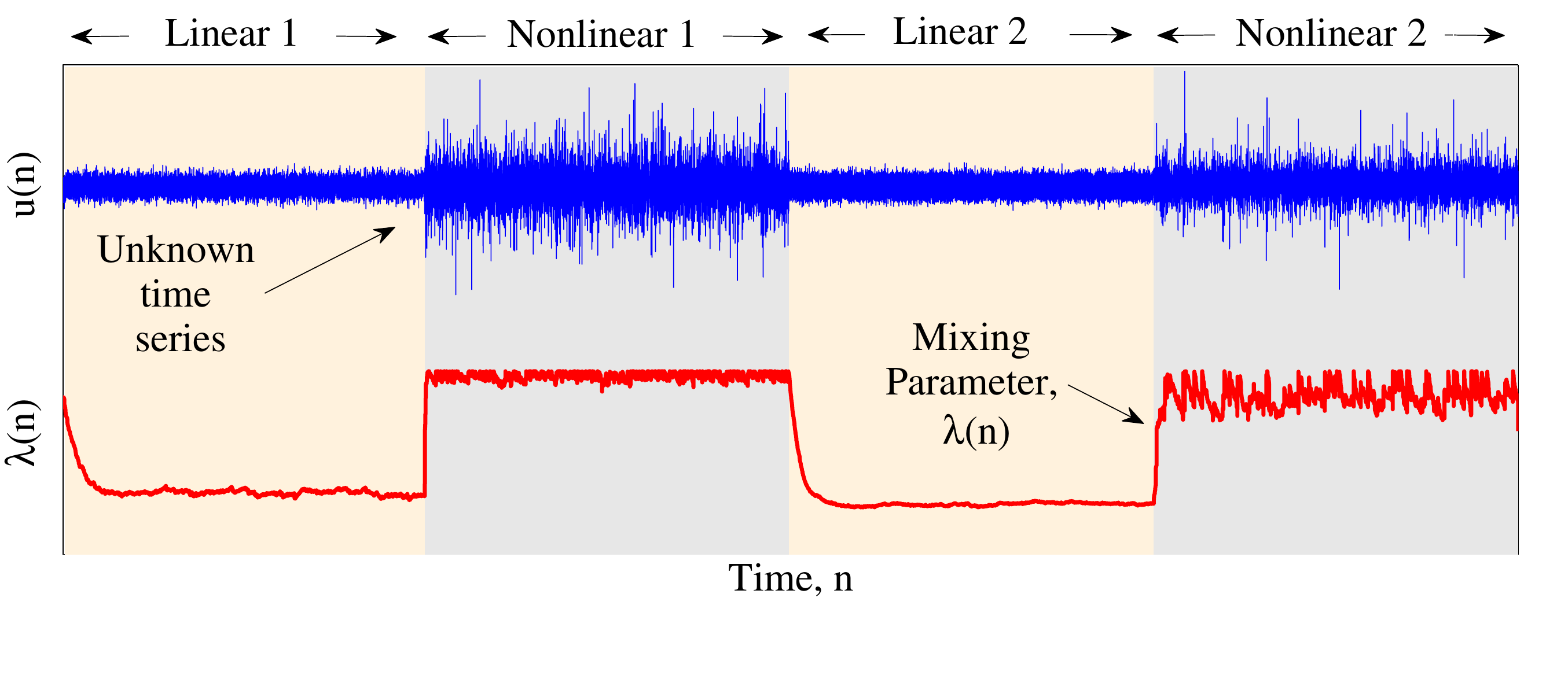}}
\caption{Signal modeling characterization using a combination of a linear and a non-linear filters. As can be seen, the mixing parameter is an effective indicator of the linear/non-linear nature of the input signal generation process.\label{modality_fig}}
\end{figure}

 As a first application example, combination of adaptive filters can be employed to characterize signal modality, for instance, whether a particular signal is generated through a linear or non-linear process. This observation has been successfully applied to track changes in the modality of different kinds of signals, including EEG \cite{Li12}, complex representations of wind and maritime radar \cite{Jelfs12b}, or speech \cite{Jelfs10}. To illustrate the idea, we consider the characterization of the linear/nonlinear nature of a signal along the lines of \cite{Jelfs10}, which uses a convex combination of a normalized LMS (NLMS) filter and a normalized nonlinear gradient descent (NNGD) algorithm to predict future values of the input signal
$$\hat{u}(n+1) = \lambda(n) \hat{u}_\text{NLMS}(n+1) + [1 - \lambda(n)] \hat{u}_\text{NNGD}(n+1).$$
The mixing parameter of the combination, $\lambda(n)$, is adapted to minimize the square error of the prediction. A value $\lambda(n) \approx 1$ means that an NLMS filter suffices to achieve a good prediction allowing us to conclude that the input signal $u(n)$ is intrinsically linear. In the other extreme, when $\lambda(n)$ approaches $0$, the predominance of the NNGD prediction suggests a non-linear input signal. Figure \ref{modality_fig} shows the time evolution of the mixing parameter for a single realization of the algorithm, when we evaluate a signal that alternates between linear autoregressive processes (of orders 1 and 4) and two benchmark nonlinear signals described in \cite[Eqs. (9)--(12)]{Jelfs10}. As it can be seen, information about the linear/nonlinear nature of the signal can be easily extracted from the evolution of the mixing parameter $\lambda(n)$.


\section{Adaptive blind equalization}
Adaptive equalizers are widely used in digital communications to remove the intersymbol interference introduced by dispersive channels. In order to avoid the transmission of pilot sequences and use the channel bandwidth in an efficient manner, these equalizers can be initially adapted using a blind algorithm and switched to a decision-directed (DD) mode after the blind equalization achieves a sufficiently low steady-state mean-square error (MSE). The selection of an appropriate switching threshold is crucial, and this selection depends on many factors such as the signal constellation, the communication channel, or the SNR.

An adaptive combination of blind and DD equalization modes can be used, where the combination layer is itself adapted using a blind criterion. This combination scheme provides an automatic mechanism for smoothly switching between the blind and DD modes. The blind algorithm mitigates intersymbol interference during the initial convergence or when abrupt changes in the channel occur. Then, when the steady-state MSE is sufficiently low, the overall equalizer automatically switches to the DD mode to get an even lower error. The main advantage of this scheme is that it avoids the need to set {\em a priori} the transition MSE level \cite{SilvaTSP2013}, and automatically changes to DD mode as soon as a sufficient equalization level is attained.

\begin{figure*}[t]
	\centering
	\includegraphics[width=13cm,trim= 5.7cm 10cm 5cm 12cm,clip=true]{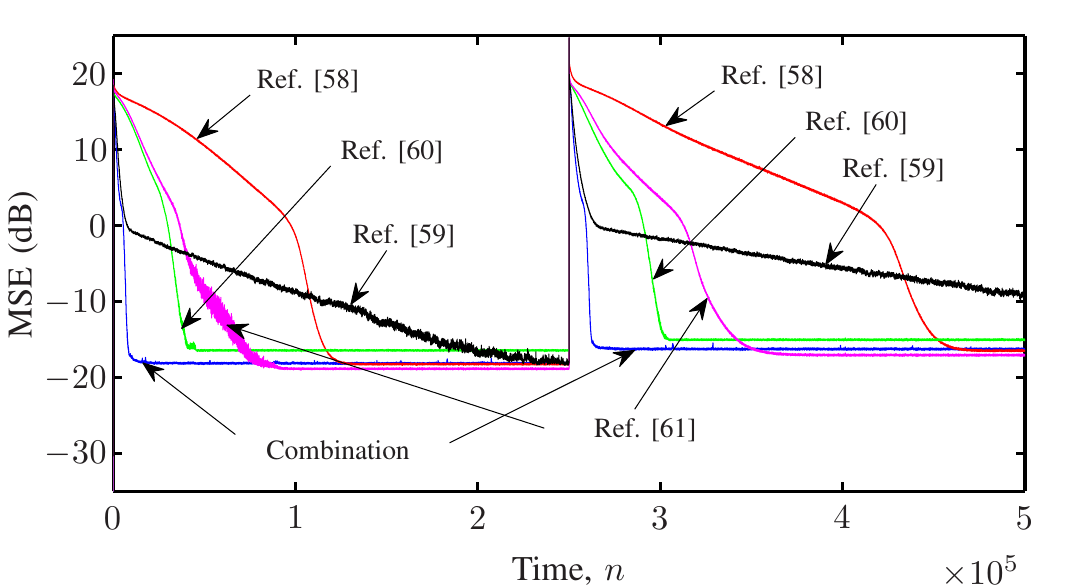}\quad~
	\includegraphics[width=13cm,trim= 4.5cm 12cm 4.5cm 12cm,clip=true]{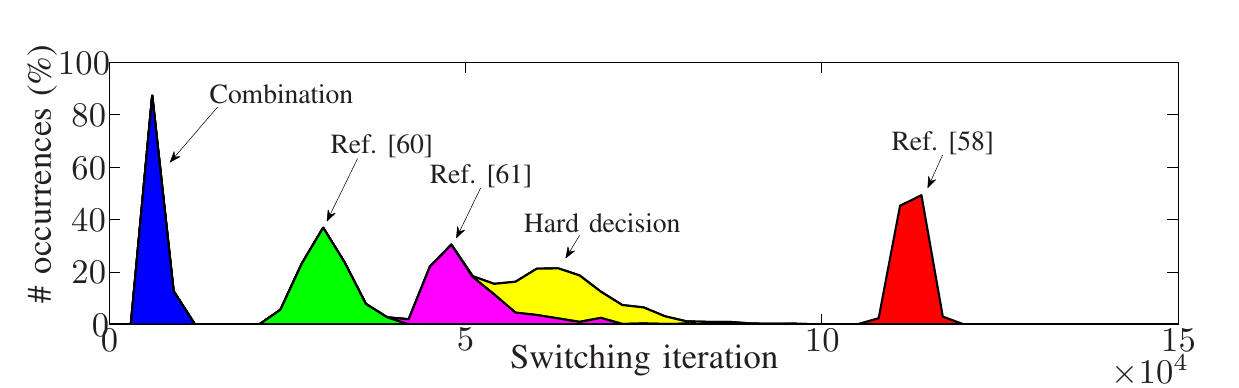}
	\caption{\label{fig:histogram} Peformance of blind equalization methods. Top panel: MSE time evolution averaged over 1000 runs. Bottom panel: Distribution of switching times from blind to DD mode (-16 dB).}
\end{figure*}
Figure \ref{fig:histogram} illustrates the performance of the combination scheme as well as of other well-known schemes for blind equalization \cite{StopGo87,Weerackody_TC_1994,Dalton,JM_Elsevier2012}. In all cases, the multimodulus algorithm (MMA) and the LMS algorithm were used for the blind and DD equalization phases, respectively. The communication scenario considers the transmission of a 256-QAM signal through a telephonic channel with an SNR of 40~dB, including an abrupt change at $n=2.5\times10^{5}$. The top panel of Fig. \ref{fig:histogram} shows the MSE evolution for all methods averaged over 1000 independent runs, whereas the bottom plot illustrates the distribution of the switching time between blind and DD modes, considering that the switching to DD mode is complete when the MSE reaches $-16$ dB for a particular run. We can conclude that the combination scheme does a better job at detecting the right moment for commuting to DD mode. Furthermore, the smaller variance observed in the switching iteration suggests that the combination method is more robust than other methods to the particularities of the simulation scenario.
 
This example shows that the combination approach can be used beyond MSE adaptive schemes. Here, we considered blind cost functions that exploit properties of high-order moments of the involved signals, both at the individual filter and at the combination layer levels. Similar uses can be conceived in other applications where high-order moments are frequently used, such as blind source separation.

\section{Sparse system identification}
Sparse systems have gathered a lot of attention. These systems are characterized by long impulse responses with only a few nonzero coefficients, and are frequently encountered in applications such as network and acoustic echo cancellation \cite{Paleo11}, compressed sensing \cite{Kalao11}, High-Definition TV \cite{Schreiber95} and wireless communication and MIMO channel estimation \cite{Guan13, Guan14}, among many others.

In the literature, several adaptive schemes have exploited such prior information to accelerate the convergence towards the sparse optimum solution. For instance, proportionate adaptive filters assign to each coefficient a different step size proportional to its amplitude \cite{Duttweiler00, Arenas09}. A recent approach \cite{Theodoridis11} promotes sparsity based on adaptive projections where the sparsity constraints are included considering $\ell_1$ balls and giving rise to a convex set whose shape is a hyperslab. An important family of sparse adaptive algorithms \cite{Hero09, Zhang14,DiLorenzo13} incorporates sparsity enforcing norms into the cost function minimized by the filter. 

The Zero-Attracting LMS (ZA-LMS) filter \cite{Hero09} uses a stochastic gradient descent rule in order to minimize a cost function that mixes the square power error and the $\ell_1$-norm penalty of the weight vector. The update equation for this adaptive scheme reads:
\begin{equation}
{\bf{w}}(n+1)={\bf{w}}(n)+\mu e(n){\bf{u}}(n) - \rho {\text{sgn}}[{\bf{w}}(n)],
\label{ZA-LMS}
\end{equation}
where function sgn[$\cdot$] extracts the sign of each element of the vector, and parameter $\rho$ controls how the sparsity is favored in the coefficient adaptation. In fact, $\rho=0$ turns \eqref{ZA-LMS} into the standard LMS update, removing any ability to promote sparsity. This scheme has shown improved performance with respect to the standard LMS algorithm when the unknown system is highly sparse; however, standard LMS outperforms ZA-LMS scheme when the system is dispersive. For this reason, there exists a tradeoff related to the degree of sparsity of the system, which unfortunately is usually unknown {\it{a priori}}, or can even vary over time. 

In this subsection, we cover two approaches specially designed to alleviate this compromise:
\begin{itemize}
\item{Scheme A: An adaptive convex combination of the ZA-NLMS filter, i.e., a sparsity-norm regularized version of the NLMS scheme and a standard NLMS algorithm \cite{Bijit14}. This approach constitutes the straightforward application of combination schemes in order to alleviate the compromise regarding the selection of parameter $\rho$ in the ZA-NLMS filter.}
\item{Scheme B: A block-wise biased adaptive filter, depicted in Fig. \ref{bias_scheme}, where the outputs of $Q$ non-overlapping blocks of coefficients of an adaptive filter are weighted by different scaling factors to obtain the overall output \cite{AzpicuetaMLSP} 
\begin{equation}
y(n)=\sum_{q=1}^{Q}{\lambda_q(n)y_q(n)}, \label{y_bias}
\end{equation}
where $y_q(n)$ is the partial output of each block, and $\lambda_q(n) \in [0,1]$ with $q=1,...,Q$ are the shrinkage factors that adapt to minimize the power of $e(n)=d(n)-y(n)$. This scheme manages the well-known bias {\it{vs}} variance compromise in a block-wise manner \cite{AzpicuetaMLSP}: if the mean-square deviation in the $q$-th block is much higher than the energy of the unknown system in this block, $\lambda_q(n)$ will tend to zero, biasing the output of the block $y_q(n)$ towards zero but reducing the output error. The shrinkage factors $\lambda_q(n)$ therefore act as estimators for the support (set of nonzero coefficients) of the filter. 
\begin{figure}
\centerline{\includegraphics[width=0.5\columnwidth,angle=0,clip=true,trim=0cm 0cm 0cm 0cm]{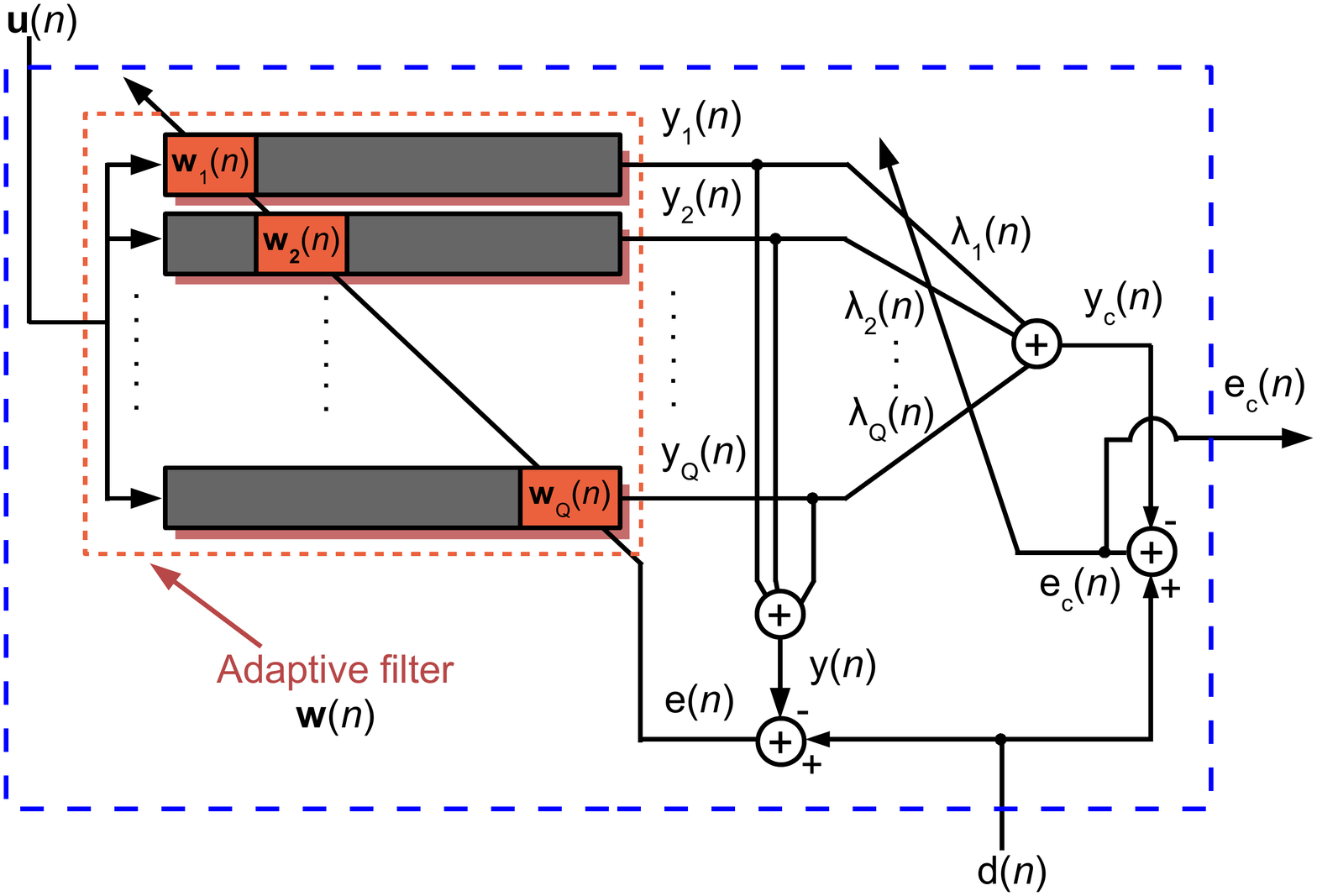}}
\caption{Block diagram of Scheme B. Each block of the adaptive filter ${\bf{w}}(n)$ is represented as a complete filter computing its output using just the indicated coefficients.} 
\label{bias_scheme}
\end{figure}

Regarding the estimation of these scaling factors $\lambda_q(n)$, with $q=1,...,Q$, it can be shown that their optimal values lie in interval $(0,1)$ \cite{AzpicuetaMLSP}. Therefore, proceeding as in \cite{LazaroTSP10}, we can reinterpret each element in the sum of \eqref{y_bias} as a convex combination between $y_q(n)$ and a virtual filter whose output remains constant and equal to zero. This is useful, because it implies that we can employ rules similar to {\em cvx-LMS} and {\em cvx-PN-LMS} reviewed in Section \ref{sec:lambda_learning} for adjusting parameters $\lambda_q(n)$. It should be noted that, for this kind of scheme, the ability to model sparse systems increases with the number of blocks $Q$, since the length of each block decreases. However, the computational cost associated with the adaptation of the mixing parameters also increases with $Q$.}
\end{itemize}

\begin{figure}
\centerline{\includegraphics[width=0.7\columnwidth,angle=0,clip=true,trim=0cm 0cm 0cm 0cm]{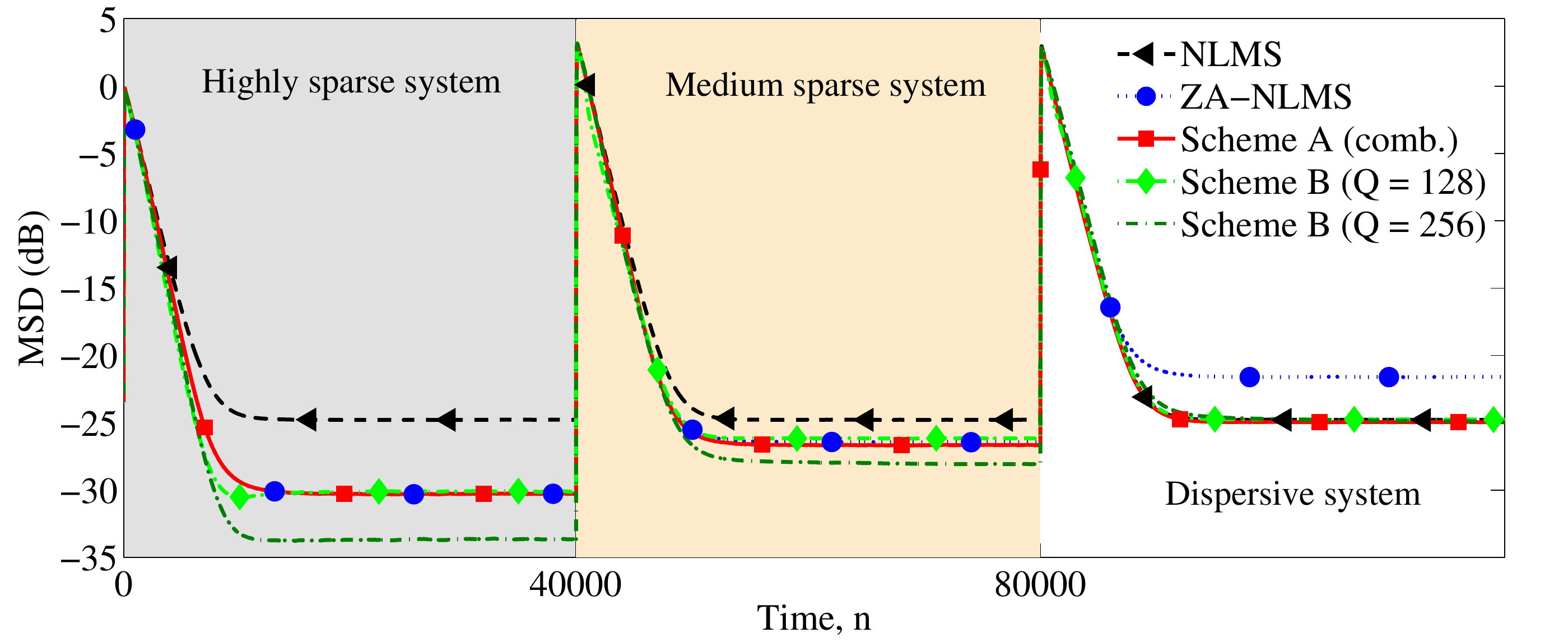}}
\caption{MSD performance of both proposed algorithms: Scheme A (and its constituent filters: An NLMS filter with $\mu=0.5$ and a ZA-NLMS with $\mu=0.5$ and $\rho =10^{-6}$) and Scheme B (based on an NLMS filter with $\mu = 0.5$) for two different configurations: $Q=128$ and $Q=256$ blocks. The sparseness of the plant varies over time.}
\label{MSD_fig}
\end{figure}

To illustrate both approaches, we have carried out a system identification experiment with an unknown plant with 1024 taps, whose sparsity degree changes over time considering white noise as input signal and SNR = 20 dB. We start with a very sparse system (only 16 active taps), that abruptly changes at $n=40000$ to a plant with 128 active coefficients, and finally, at $n=80000$, we employ a more dispersive unknown system with 512 active taps. Figure \ref{MSD_fig} shows the MSD reached by Scheme A and Scheme B, considering two different possibilities for the number of blocks $Q$.

As expected, Scheme A shows a robust behavior with respect to the sparsity degree of the unknown plant, behaving as well as its best component filter. Scheme B provides an attractive alternative, whose performance improves when the length of each block is reduced. Comparing both schemes, the computational cost of Scheme B is smaller than that of Scheme A, even when $Q=256$. In this situation, we also observe that Scheme B outperforms Scheme A in terms of MSD for highly and medium sparse systems.

\section{Acoustic echo cancellation}\label{sec:acoustic.echo}
Combination filters have been successfully employed in acoustic signal processing applications, such as active noise cancellation \cite{Ferrer13,George14}, adaptive beamforming \cite{Comminiello13}, and acoustic echo cancellation, where combination schemes have been used in linear cancellation (considering both time-domain \cite{Arenas09} and frequency-domain \cite{Ni10} schemes), and in nonlinear acoustic echo cancellation \cite{Azpicueta11,Azpicueta13,Comminiello13b}. 

\begin{figure}[t]
\centerline{\includegraphics[width=0.5\columnwidth,angle=0,clip=true,trim=2.2cm 3cm 5cm 5cm]{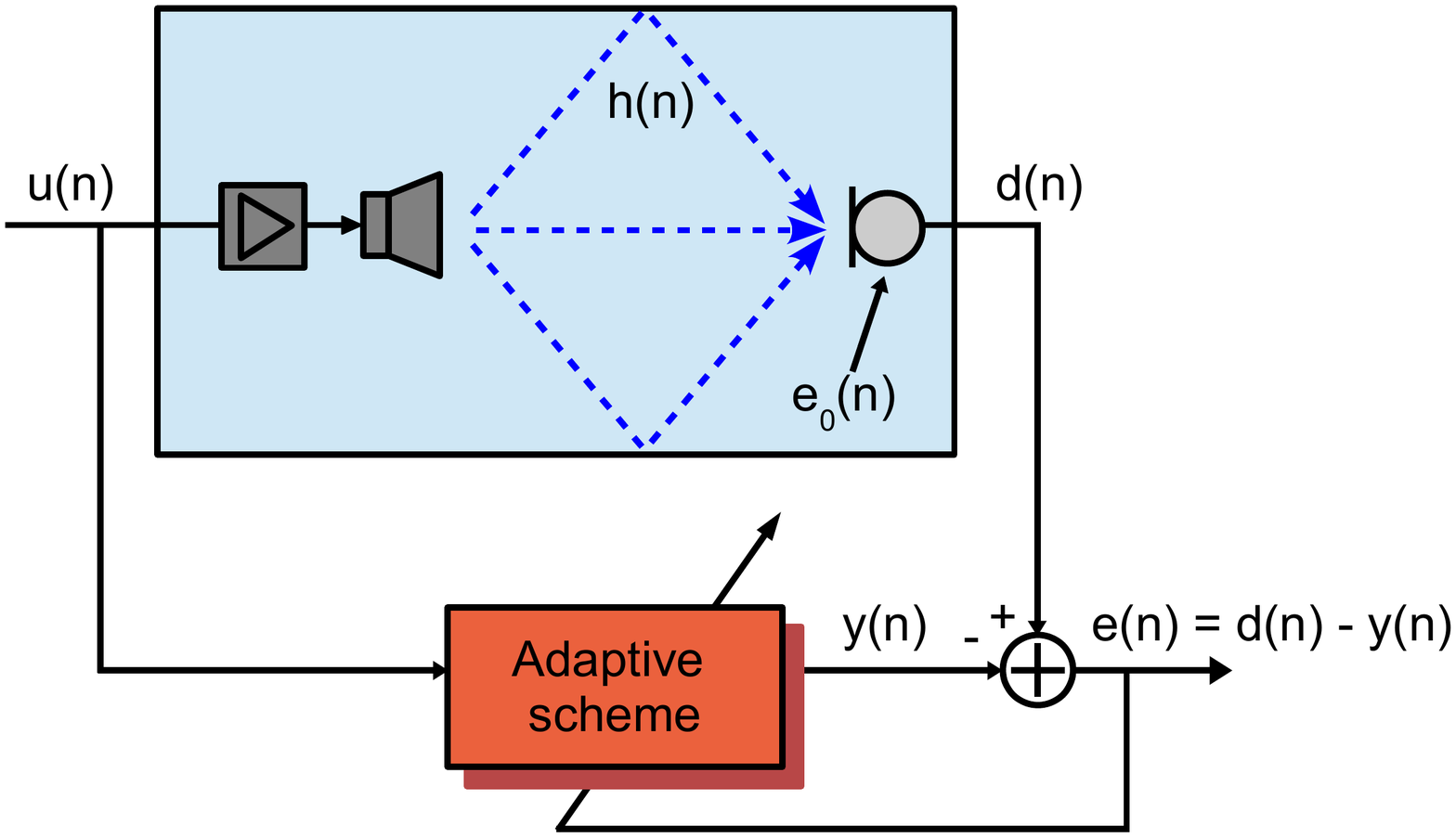}}
\caption{Acoustic echo cancellation scenario. Signal ${\bf{u}}(n)$ stands for the input signal, $e_0(n)$ is the background noise at microphone location, $h(n)$ represents the room impulse response that describes the acoustic propagation (linear in nature), and where depending on the behavior of the loudspeaker and the power amplifier, nonlinear distortion can be found in desired signal $d(n)$. The adaptive scheme tries to minimize the power of the error cancellation $e(n)$, incorporating nonlinear models if necessary.}
\label{Echo}
\end{figure}

The abundance of portable devices has been motivating the inclusion of nonlinear models in the adaptive scheme of acoustic echo cancellers (see Fig. \ref{Echo}), in order to compensate for the nonlinear distortion caused by low-cost loudspeakers fed by power amplifiers driven at high levels. One of the main solutions used for this purpose is the Volterra filter (VF), which is able to represent a large class of nonlinearities with memory \cite{Mathews91}.

The output of a VF can be calculated as the addition of partial outputs of different kernels, each one representing an order of nonlinearity, i.e., 
\begin{equation}
y(n)= \sum_{p=1}^P{y_p(n)}= \sum_{p=1}^P{\sum_{i_1=0}^{N_p-1}\cdots{\sum_{i_p=i_p-1}^{N_p-1}h_{p,i_1,...,i_{p}}\prod_{q=1}^{p}{u(n-i_q)}}},
\label{VF}
\end{equation}
where $y_p(n)$ corresponds with the output of the $p$-th kernel with length $N_p$, $P$ is the number of kernels that compose the VF and $h_{p,i_1,...,i_{p}}$ represents the adaptive coefficients of the $p$-th kernel. For instance, the output of a VF of order $P=2$ can be obtained as $y(n) = y_1(n) + y_2(n)$, where $y_1(n)$ is the output of the linear kernel, and $y_2(n)$ represents the output of the quadratic kernel, responsible for the modeling of second-order nonlinearities. Despite the large computational cost associated with the operation of a VF, the success of this nonlinear filter lies in the simplicity of the adaptation of the coefficients in each kernel $h_{p,i_1,...,i_{p}}$, which permits the employment of update rules similar to those of linear adaptive filters. For this reason, VFs present similar tradeoffs to those of linear filters, in particular the selection of the step size, and the memory of the kernels.  

\begin{figure}
\centerline{\includegraphics[width=1.1\columnwidth,angle=0,clip=true,trim=0cm 0cm 0cm 0cm]{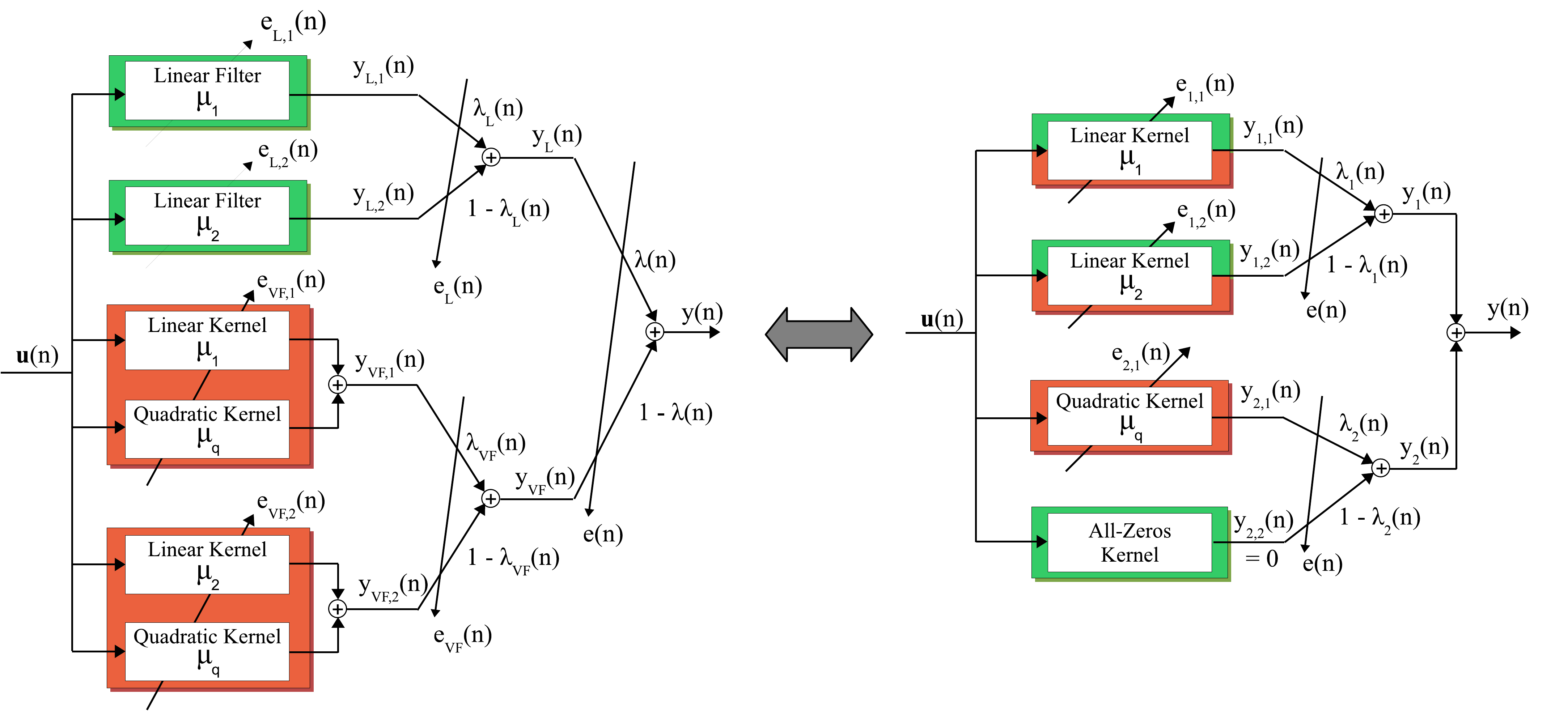}}
\caption{Nonlinear acoustic echo canceller based on combination of filters. On the left: A hierarchical combination of filters (linear and VFs) where each filter adapts employing its own error signal, for instance $e_{{\text{L}},1}(n) = d(n)-y_{{\text{L}},1}(n)$. On the right: A special VF based on the CK, where each kernel adapts using an error signal employing its own output and the combined outputs of the kernels of different orders (see \cite{Azpicueta11}). The mixing parameters $\lambda_1(n)$ and $\lambda_2(n)$ update to minimize the power of the global error signal, i.e., $e(n)=d(n)-y(n)$.}
\label{CK_Scheme}
\end{figure}

In \cite{Azpicueta11}, two different schemes were proposed to alleviate such compromises, named combination of Volterra filters and combination of kernels (CK) ---see Fig. \ref{CK_Scheme}. The first approach constitutes a straightforward application of the combination idea, i.e., the output of the combined scheme is calculated mixing different VFs outputs, each one obtained by means of \eqref{VF}, and following any of the schemes included in Section VI. However, the second  scheme proposes a special kind of VF, where each kernel is replaced by a combination of kernels with complementary settings. For the case of a combination of two kernels per order, the output of the CK scheme reads
\begin{equation}
y(n)= \sum_{p=1}^P{y_p(n)}= \sum_{p=1}^P{\lambda_p(n)y_{p,1}(n)+[1-\lambda_p(n)]y_{p,2}(n)},
\label{C_Ks}
\end{equation}
where the outputs of two kernels of order $p$, $y_{p,1}(n)$ and  $y_{p,2}(n)$, are combined by means of the mixing parameter $\lambda_p(n)$. Both algorithms reach similar performance but CK is a more attractive scheme regarding the computational cost \cite{Azpicueta11}.

We should notice that the performance of VFs is also subject to a tradeoff related to the number of kernels, since the degree of nonlinearity present in the filtering scenario is rarely known {\it{a priori}}. This compromise directly affects the nonlinear acoustic echo cancellers: if only linear distortion is present in the echo path, the adaptation of nonlinear kernels degrades the performance of the whole VF, which behaves worse than a simple linear adaptive filter.

In order to deal with these compromises, a nonlinear acoustic echo canceller, represented in Figure \ref{CK_Scheme} using both the combination of Volterra filters and the CK approaches, was proposed in \cite{Azpicueta11} considering two requirements: the robustness with respect to the presence or absence of nonlinear distortion, and the fact that the acoustic room impulse response $h(n)$ is strongly variant. The first requirement is dealt with considering a combination of a quadratic kernel and a {\it{virtual}} kernel named {\it{All-Zeros Kernel}}, with all coefficients equal to zero and no adaptation, whose partial output is always zero. Regarding the second requisite, a combination of two linear kernels with different step sizes serves to provide a fast reconvergence and a low steady-state error. 

The output of this algorithm, considering the CK scheme, is:
\begin{equation}
y(n)=\underbrace{\lambda_1(n)y_{1,1}(n)+[1-\lambda_1(n)]y_{1,2}(n)}_{\text{linear part}}+\underbrace{\lambda_2(n)y_{2,1}(n)+[1-\lambda_2(n)]\cdot0}_{\text{nonlinear part}},
\label{y_out}
\end{equation}
where $y_{1,1}(n)$ and $y_{1,2}(n)$ are the outputs of two linear kernels with different step sizes, $\lambda_1(n)$ is the mixing parameter to combine them, $y_{2,1}(n)$ is the output of the quadratic kernel, and $\lambda_2(n)$ is the mixing parameter to weight the estimation of the nonlinear echo.


\begin{figure}
\centerline{\includegraphics[width=.7\columnwidth,angle=0,clip=true,trim=0cm 0 0cm 0]{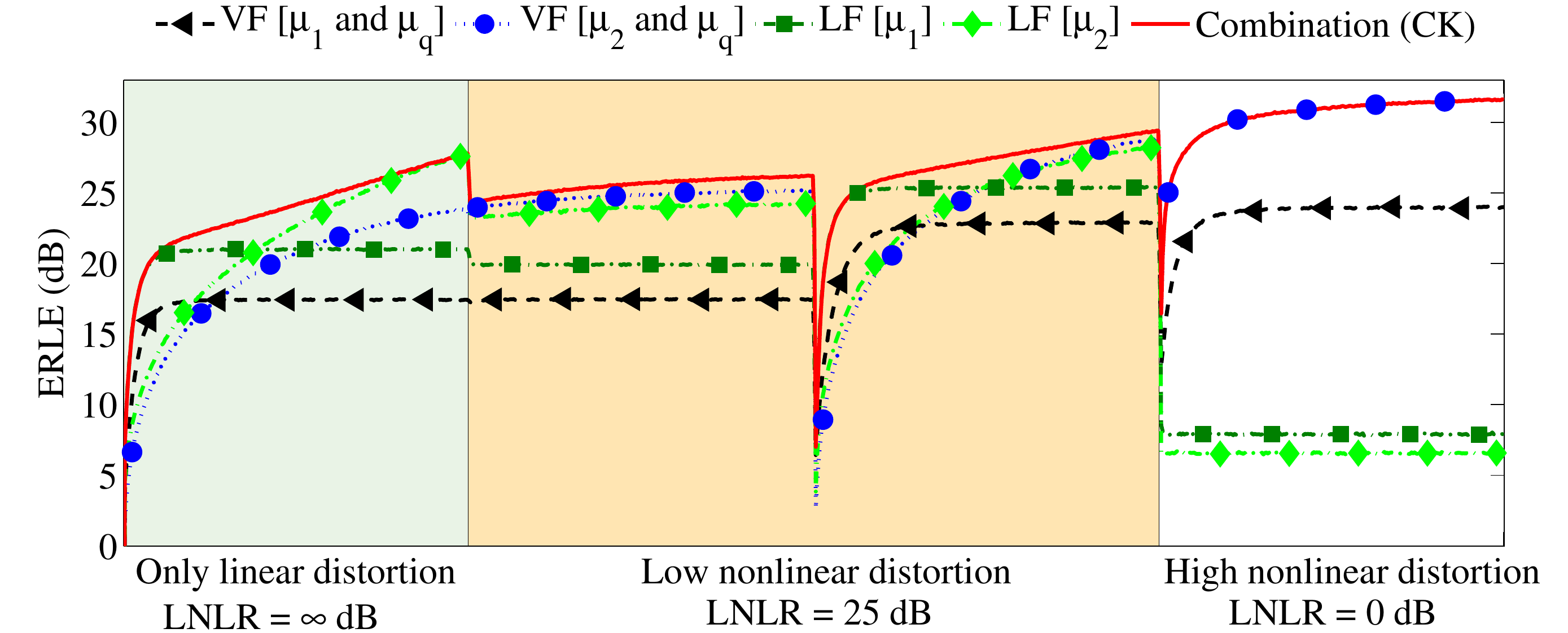}}
\centerline{\includegraphics[width=.7\columnwidth,angle=0,clip=true,trim=0cm 0 0cm 0]{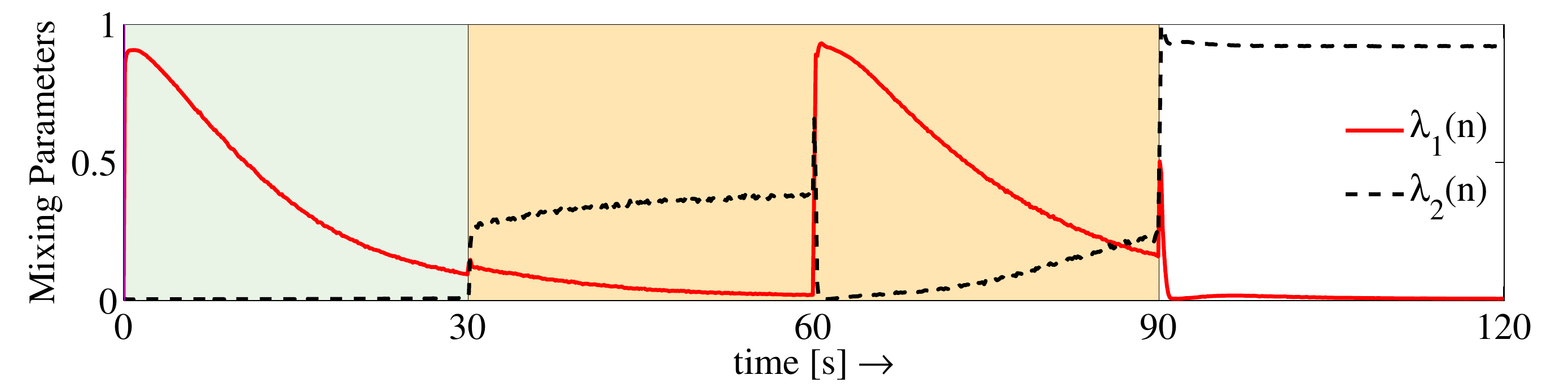}}
\caption{Performance of the proposed scheme as a function of the linear-to-nonlinear ratio of echo powers (LNLR), including an abrupt change in the room impulse response at $t=60$ s. From top to bottom: ERLE achieved by the combination of kernels scheme and of all possible filters (linear and VFs) using its constituent kernels; time evolution of the mixing parameters. In this experiment, we employ Laplacian speech-like colored noise as input signal; experiments with real speech can be found in \cite{Azpicueta11}.}
\label{ERLEfig}
\end{figure}

Figure \ref{ERLEfig} represents the performance of the proposed nonlinear acoustic echo canceller, compared with that of all the adaptive filters (linear and nonlinear) that could be designed with its kernels, in terms of the Echo Return Loss Enhancement (ERLE) computed as:
\begin{equation}
\text{ERLE}(n) \triangleq 10\text{log}\frac{\mathbb{E}\{[d(n)-e_0(n)]^2\}}{\mathbb{E}\{[e(n)-e_0(n)]^2\}}.
\label{ERLE}
\end{equation}
As it can be seen, the proposed scheme reaches the best performance, behaving as a combination of linear filters when only linear distortion is present ($t\leq 30$ s). In this case, $\lambda_2(n) \approx 0$ and the quadratic kernel is not considered in the output of the filter, without degrading the performance of the proposed scheme with respect to that of linear filters. At the end of the experiment ($t > 90$ s), when strong nonlinear distortion is considered,  $\lambda_2(n) \approx 1$, and the output of the quadratic kernel is fully incorporated, providing the ability to model nonlinearities. In an intermediate case, during 30 s $< t \leq 90$  s, the proposed scheme slightly outperforms both linear and nonlinear filters. In addition, the algorithm shows a suitable reconvergence when the room impulse response abruptly changes, thanks to the combination of two linear kernels with different step sizes, as it can be seen after $t = 60$ s.

\section{Conclusions and open problems}
Combinations of adaptive filters constitute a powerful approach to improve the performance of adaptive filters. In this paper, we have reviewed some of the most popular combination schemes, highlighting their theoretical properties and limits. Practical algorithms to combine adaptive filters need to implement estimation methods to adjust the combination layer, taking into account the possibly time-varying conditions in which the filter operates.

We have reviewed several of the methods that have been proposed in the literature, paying special attention to gradient methods. Power-normalized algorithms are particularly interesting, since they simplify the selection of parameters and result in a more robust behavior when the statistics of the filtering scenario are (partly) unknown, which is frequently the case.
The versatility of the approach has been demonstrated through several examples in a variety of applications. We have seen that, in all studied scenarios, combination schemes offer competitive performance when compared to state-of-the-art methods for each application. This fact, together with the inherent simplicity of the approach, make combination structures attractive for demanding applications requiring enhanced performance, as illustrated by the several examples and references mentioned in the body of the article. 

In our opinion, some of the most interesting open problems to be addressed are:
\begin{itemize}
\item the selection and design of component filters with reduced cross-EMSE, with the goal to minimize the overall EMSE,
\item providing and exploiting strategies to reduce the cost of combined schemes, trying to develop new structures whose complexity is close to that of an individual filter, and 
\item extensions to other application domains where different kinds of compromises and performance tradeoffs may be present.
\end{itemize}

\section*{Acknowledgments}

The work of Arenas-García and Azpicueta-Ruiz was partially supported by the Spanish Ministry of Economy and Competitiveness (MINECO) under projects TEC2011-22480 and PRI-PIBIN-2011-1266. The work of Silva was partially supported by CNPq under Grant 304275/2014-0 and by FAPESP under Grant 2012/24835-1.  The work of Vítor H. Nascimento was partially supported by CNPq under Grant 306268/2014-0 and FAPESP under grant 2014/04256-2.
The work of A. H. Sayed was supported in part by NSF grants CCF-1011918 and ECCS-1407712.

We are grateful to the colleagues with whom we have shared discussions and co-authorship of papers along these research lines, especially Prof. Aníbal R. Figueiras-Vidal.

%
%

\bibliographystyle{IEEEbib}

\end{document}